\documentclass[aps,twocolumn,a4paper,showpacs]{revtex4}
\usepackage{graphicx}
\usepackage{amsmath,amssymb}
\usepackage{enumerate}
\usepackage{subfigure}
\usepackage{tabularx}

\begin{document}
\title{Orthonormal wave functions for periodic fermionic states under an applied magnetic Field}
\author{Edinardo I. B. Rodrigues}
\affiliation{Universidade Federal Rural de Pernambuco,54518-430,Cabo de Santo Agostinho, Pernambuco, Brazil}
\author{Mauro M. Doria}
\affiliation{Instituto de F\'{\i}sica, Universidade Federal do Rio de Janeiro, 21941-972 Rio de Janeiro, Brazil}%

\begin{abstract}
We report an  infinite number of orthonormal wave functions bases for the quantum problem of a free particle in presence of an applied external  magnetic field.
Each set of  orthonormal wave functions (basis) is labeled by an integer $p$, which is the number of magnetic fluxons trapped in the unit cell.
These bases are suitable to describe particles whose  probability density is periodic and defines a lattice in position space.
The present bases of  orthonormal wave functions unveils fractional effects since the number of particles in the unit cell is independent of the number of trapped fluxons.
For a single particle under $p$ fluxes in the unit cell, and confined to the lowest Landau level,
the probability density vanishes in $p$ points, thus each zero is  associated to a fraction $1/p$ of the particle.
Remarkably the case  of $n+1$ filled Landau levels, hence with a total of $N=(n+1)p$ fermions, $n$ being the highest filled Landau level, the  density displays an egg-box pattern with $p^2$ maxima (minima) which means that a $(n+1)/p$ fraction of flux is associated to every one of these  maxima (minima).
We also consider the case of particles interacting through the magnetic field energy created by their own motion and find an attractive interaction among them in case they are confined to the lowest Landau level ($n=0$).
The well-known de Haas-van Alphen oscillations are retrieved within the present  orthonormal basis of wave functions  thus providing evidence of its correctness.
\end{abstract}
\date{\today}
\maketitle
\section{Introduction}

The quantum problem of free fermions in a magnetic field has been long investigated since the early days of Quantum Mechanics, and its study is nowadays part of the educational training of any physics student~\cite{flugge47,kittel76,ashcroft76,rohlf94,band13}.
Although nearly ninety years have passed since Landau firstly explained the diamagnetism of metals~\cite{landau30}, based on a free electron model, the study of particles in a magnetic field has remained a subject of interest as new and startling phenomena are still unfolding from it~\cite{barrier20}, such as
the quantum Hall effect~\cite{doucot05,vonklitzing20}.
In this paper we report novel and interesting properties in case that fermions form a lattice state, and so, the probability density is periodic in position space.
Our results stem from the finding of new bases of orthonormal wave functions, which are reported here and follow from  Abrikosov`s solution for the vortex lattice~\cite{abrikosov56}
The Landau gauge is used and a rectangular  unit cell is taken with dimension $L_1$ and $L_2$.
The orthonormal wave function basis  features $p$ magnetic fluxons trapped in the unit cell ($\Phi=p\Phi_0$, $\Phi_0 \equiv hc/e$, $e$ the electronic charge) and the choice of $p$ fixes a distinct set of orthonormal functions.
Hence in the present formalism each Landau level, labeled by $n$ has $p$ available states, each one associated to a distinct wave function that belongs to this orthonormal set.
For simplicity we ignore the spin degree of freedom such that each level fits a single particle.
The present approach allows for the treatment of situations such that  the number of particles and the number of fluxons in the unit cell are not necessarily the same, thus potentially useful for cases with fractional charge and flux.
We treat a few cases in details to exemplify the fractional effects.
For instance the case of the full lowest Landau level, with $p$ particles which is the same as the number of fluxons,  shown in Fig.~\ref{fig1}.
There one sees the presence of $p^2$ maxima (minima) of the probability density, thus there is $1/p$ fluxons associated to each maxima (minima),  as shown in Table~\ref{tab-ratio}.
The case of a single particle in a unit cell with $p$ fluxons is seen in Fig.~\ref{fig2}.
The probability density has $p$ zeros and so, each zero is associated with $1/p$ of a particle.
Finally in Fig.~\ref{fig3} is the case of $n+1$ fully filled Landau levels, and so with $N=(n+1)p$ particles or fluxons.
We find the remarkable property that the probability density displays $p^2$ maxima (minima) in  position space within the unit cell, similarly to an egg-box arrangement, which is  Fig.~\ref{fig1} result  extended to higher Landau levels.
Consequently the number of particles per maximum is fractional, and given by $(n+1)/p$, as shown in Table~\ref{tab-ratio}.
The above examples correspond to free fermions in a magnetic field.
We also treat in this paper the curious case that the local magnetic field created by the motion of the fermions is also taken into account, provided that the particles are confined to the first Landau level ($n=0$).
We report here that the magnetic energy of this system is negative thus causing an attraction between the fermions.
For this special case the so-called first order equations~\cite{gomes16} apply.
They were firstly used by A.A. Abrikosov to discover vortices in superconductors~\cite{abrikosov56}.\\

To show the correctness of the present approach we use the present formalism to  the de Haas-Van Alphen effect and retrieve some well-known properties, such as  the periodicity of the energy with respect to the applied field $H_3$, and also with respect to  $1/H_3$, that allows for the measurement of the Fermi surface area.
The de Haas-van Alphen effect is known to reveal quantum oscillatory phenomena in
metals that unveils fundamental properties directly obtained from the magnetisation $M$, which  is a thermodynamic function of state~\cite{kittel76,ashcroft76,rohlf94,band13}.
This means that theoretical models for the Fermi surface can be checked in a very rigorous manner through the de Haas-van Alphen effect~\cite{harrison96,lukyanchuk11}.
%
\section{Periodic solutions for the  Schr\"odinger equation of a free particle in a magnetic field}
In this section we obtain  solutions of the  Schr\"odinger for free particles in presence of a magnetic field $H_3$, under the condition of spatial periodicity.
The Hamiltonian is well-known and given by,
\begin{equation}
\frac{1}{2m}\vec{D}^{2}\psi=E\psi,
\label{eq0:01}
\end{equation}
where $\vec{D}=\hat{x}_{1}D_{1}+\hat{x}_{2}D_{2}+\hat{x}_{3}D_{3}$. The covariant derivative is $D_{j} \equiv -i\hbar \nabla\!_{j} -(e/c)A_{j}$, $j=1,\,2$ e $3$, $e$ is the particles charge and  $A_{j}$ is the vector potential.
The magnetic field is set along the direction $\hat{x}_3$ such that the Landau gauge is given by $\vec{A}=(-H_{3}x_{2},0,0)$, $\vec{\nabla}\times\vec{A}=\hat{x}_{3}H_{3}$.
Uniaxial symmetry is assumed and so, there is no derivative along $\hat{x}_3$  since $A_{3}=0$.
Only the derivatives $D_{1}$ e $D_{2}$ remain and so, the index is limited to  $j=1$ e $2$.
The wave function is described by the coordinates perpendicular to the direction of the applied field, $\psi=\psi(x_{1},x_{2})$.\\

Assume a rectangular unit cell, with dimensions $L_1$ e $L_2$ in this plane $(x_{1},x_{2})$.
We seek states that are periodic on this lattice and for this purpose make demands on the wave function under $x_{1}\rightarrow x_{1}+L_{1}$ and  $x_{2}\rightarrow x_{2}+L_{2}$ to guarantee that $|\psi(x_{1},x_{2})|^{2}$ be periodic.
This means to impose quasi-periodicity along the coordinate $x_1$,
\begin{equation} \label{quasi1}
\psi(x_{1}+L_{1},x_{2})=e^{i\eta_1}\psi(x_{1},x_{2}),
\end{equation}
as the phase  $e^{i\eta_1}$ does not affect the periodicity condition $|\psi(x_{1}+L_{1},x_{2})|^{2} = |\psi(x_{1},x_{2})|^{2}$.
Similarly the periodicity along the coordinate $x_2$, $|\psi(x_{1},x_{2}+L_{2})|^{2} = |\psi(x_{1},x_{2})|^{2}$, is a consequence of the demand that,
\begin{equation} \label{quasi2}
\psi(x_{1},x_{2}+L_{2})=e^{i\eta_2}\psi(x_{1},x_{2}),
\end{equation}
where $e^{i\eta_2}$ is an arbitrary phase.\\

The general solution of Eq.~\eqref{eq0:01} in the Landau gauge is well known to be given by $\psi\equiv \psi_{n,k}(x_{1},x_{2})=e^{ikx_{1}}f_{n}(x_{2})$.
Thus along one of the coordinates, $x_1$, is a plane wave, whereas along the other, $x_2$, is a harmonic oscillator.
Therefore there are two quantum indices, $k$ and $n$, the latter index defines the Landau level.
Hence Eq.~\eqref{eq0:01} acquires the following form.
\begin{equation}
\!\! \left[-\frac{\hbar^2}{2m}\frac{\partial^2}{\partial x^{2}_{2}} + \frac{1}{2}m\omega^{2}_{c}\left(x_{2}- x'_{2} \right)^{2} \right]f_{n}(x_{2}) = E_{n} f_{n}(x_{2}),
\label{eq0:03}
\end{equation}
where $x'_{2}= - (\hbar ck)/(eH_{3})$ and $\omega_{c}$ is the Larmor frequency given by,
\begin{equation}
\omega_{c}=\frac{eH_{3}}{mc}.
\label{eq0:04}
\end{equation}
The analogy of Eq.~\eqref{eq0:03} with the harmonic oscillator yields the eigenvalues as,
\begin{equation}
E_n=\hbar\omega_{c}\left(n+\frac{1}{2}\right), \quad n=0,1,2,3...,
\label{eq0:05}
\end{equation}
and shows that the Landau levels are equally spaced in energy and separated by $\hbar\omega_{c}$.

The solution for $f_{n}(x_{2})$ is given by, %
\begin{equation}
f_{n}(x_{2})= A_{n}H_{n}(\bar{x}_{2})e^{-\frac{1}{2}\bar{x}^{2}_{2}},
\label{eq0:06}
\end{equation}
where  $A_{n}$ is a constant to be determined, $H_{n}(\bar{x}_{2})$ are the Hermite polynomials and the variable $\bar{x}_{2}$
is defined by,
\begin{eqnarray}
\bar{x}_{2} \equiv \sqrt{\frac{eH_{3}}{\hbar c}}\left( x_{2} + \frac{c\hbar}{eH_{3}}k \right).
\label{eq0:07}
\end{eqnarray}
Hence the wave function is given by,
\begin{equation}
\psi_{n,k}(x_{1},x_{2})=A_{n}e^{ikx_{1}}H_{n}(\bar{x}_{2})e^{-\frac{1}{2}\bar{x}^{2}_{2}}.
\label{eq0:08}
\end{equation}
The wave functions  $\psi_{n,k}(x_{1},x_{2})$, Eq.~\eqref{eq0:08}, are orthogonal, and satisfy the following condition.
\begin{equation}
\int d^{2}x \; \psi^{*}_{m,k'}\psi_{n,k}=\delta(k-k')\delta_{nm}.
\label{eq0:09}
\end{equation}
This is easily checked, by taking $\psi_{m,k'}$ e $\psi_{n,k}$, and computing the integral,
\begin{equation}
\int d^{2}x \; \psi^{*}_{m,k'}\psi_{n,k}= A_{m}A_{n}I_{x_1}I_{x_2},
\label{eq0:10}
\end{equation}
where
\begin{eqnarray}
I_{x_1}&=&\int^{+\infty}_{-\infty} dx_{1} \; e^{i(k-k')x_{1}},
\label{eq0:11a}
\end{eqnarray}
and
\begin{eqnarray}
I_{x_2}&=&\int^{+\infty}_{-\infty} dx_{2}\; H_{m}(\bar{x}'_{2})H_{n}(\bar{x}_{2})e^{-\frac{1}{2}(\bar{x}'^{2}_{2}+\bar{x}^{2}_{2})},
\label{eq0:11b}
\end{eqnarray}
such that  $\bar{x}_{2}$ is given by Eq.~\eqref{eq0:07} and $\bar{x}'_{2}$ is equivalent to Eq.~\eqref{eq0:07} for $k'$.
The integration in  $x_1$ yields that $I_{x_1}=2\pi\delta(k-k')$.
Then $\bar{x}'_{2}=\bar{x}_{2}$ and deriving both sides of Eq.~\eqref{eq0:07} gives that
$dx_{2}=\sqrt{\frac{\hbar c}{eH_{3}}}d\bar{x}_{2}$.
Hence $I_{x_2}=\sqrt{\frac{\hbar c}{eH_{3}}}\int^{\infty}_{-\infty} d\bar{x}_{2}H_{m}(\bar{x}_{2})H_{n}(\bar{x}_{2})e^{-\bar{x}^{2}_{2}}$.
One obtains that $I_{x_2}=0$ if $m\neq n$ and $I_{x_2}=\sqrt{\frac{\hbar c}{eH_{3}}}2^{n}n!\sqrt{\pi}$ if $m=n$.
Combining these results in Eq.~\eqref{eq0:10} gives the constant of Eq.~\eqref{eq0:08}.
\begin{equation}
A_{n}=\left( \frac{eH_{3}}{4\pi^{2}\hbar c} \right)^{1/4}\left(2^{n}n!\sqrt{\pi} \right)^{-1/2}.
\label{eq0:12}
\end{equation}
So far the orthonormality of the wave functions, expressed in Eq.~\eqref{eq0:09} is not limited to the unit cell, and this is where news bases can be introduced as their wave functions are orthonormal in a unit cell.
\\

In power of the well-known solution for the Schr\"odinger equation, given by Eq.~\eqref{eq0:08}, we revisit the periodicity to unveil startling properties.
Along  $x_1$ the quasi-periodicity is easily checked:
$\psi_n(x_{1}+L_{1},x_{2})=e^{ikL_{1}}\psi_n(x_{1},x_{2})=A_{n}e^{ikL_{1}}\sum_{k} c_{n,k}e^{ikx_{1}}H_{n}(\bar{x}_{2})e^{-\frac{1}{2}\bar{x}^{2}_{2}}$, and
$e^{i\eta_1}=e^{ikL_{1}}$.
However the quasi periodicity along $x_2$ is only possible for the wavefunction  summed over $k$,
\begin{equation}
\psi_n(x_{1},x_{2})=\sum_{k} c_{n,k}\psi_{n,k}(x_{1},x_{2}),
\label{eq0:02}
\end{equation}
which introduces the so far free parameters, $c_{n,k}$.
However the above wave function spoils the quasi periodicity along $x_1$, unless if the sum over $k$ is limited to the following values,
\begin{eqnarray}
k=\frac{2\pi}{L_1}l, \quad l=0, \pm 1, \pm 2,...,
\label{eq0:13}
\end{eqnarray}
In this case, $e^{i\eta_1}=1$, and the coefficients belong to the discrete set  $c_{n,k} = c_{n,l}$.
A remarkable consequence of the quasi periodicity condition along $x_2$  is the quantization of the magnetic flux in the unit cell.
There should be an integer number of $p$ magnetic flux inside the unit cell,
\begin{eqnarray}
p\Phi_{0} = H_{3}L_{1}L_{2},
\label{eq0:14}
\end{eqnarray}
where  $\Phi_0$ is the unit flux defined by a single electronic charge,
\begin{eqnarray}
\Phi_{0} = \frac{hc}{e}=4,14\cdot 10^{-7}G\cdot \text{cm}^{2}.
\label{eq0:15}
\end{eqnarray}
to have quasi periodicity along $x_2$.
This  also limits the number of coefficients in Eq.~\eqref{eq0:02} to just $p$ free coefficients, since it must hold that,
\begin{eqnarray}
c_{n,l+p}=c_{n,l}.
\label{eq0:18}
\end{eqnarray}
To check this write the wave function as,
\begin{eqnarray}
\psi_n(x_{1},x_{2})=\sum_{l} c_{n,l}\psi_{n,l}(x_{1},x_{2}),
\label{eq0:16}
\end{eqnarray}
where
\begin{eqnarray}
\psi_{n,l}(x_{1},x_{2})=A_{n}e^{i\frac{2\pi}{L_{1}}lx_{1}}H_{n}\left(g_{l}(x_{2})\right)e^{-\frac{1}{2}\left( g_{l}(x_{2})\right)^{2}}
\label{eq0:17}
\end{eqnarray}
with $g_{l}(x_{2})\equiv\sqrt{\frac{2\pi H_{3}}{\Phi_{0}}}\left( x_{2} + \frac{\Phi_{0}}{H_{3}L_{1}}l\right)$, and $\Phi_{0}$  given by \eqref{eq0:15}.
To calculate $\psi_n(x_{1},x_{2}+L_{2})$, firstly notice that under the flux quantization condition one obtains that, $L_{2}+ (\Phi_{0}/H_{3}L_{1})l = (\Phi_{0}/H_{3}L_{1})(l+p)$.
Next write $e^{i\frac{2\pi}{L_{1}}lx_{1}}$ as $e^{-i\frac{2\pi}{L_{1}}px_{1}}e^{i\frac{2\pi}{L_{1}}(l+p)x_{1}}$, to obtain that
$\psi_n(x_{1},x_{2}+L_{2})=e^{-i\frac{2\pi}{L_{1}}px_{1}}A_{n}\sum_{l} c^{n}_{l+p}e^{i\frac{2\pi}{L_{1}}(l+p)x_{1}}H_{n}(\bar{g_{l}}(x_{2}))e^{-\frac{1}{2}\left( \bar{g_{l}}(x_{2})\right)^{2}}$ com $\bar{g_{l}}(x_{2})=\sqrt{\frac{2\pi p}{L_{1}L_{2}}}\left( x_{2} + \frac{L_{2}}{p}(l+p)\right)$.
Define  $l'=l+p$ to retrieve the original form, namely,  $\psi_n(x_{1},x_{2}+L_{2})=e^{-i\frac{2\pi}{L_{1}}px_{1}}\psi_n(x_{1},x_{2})$. Thus it holds that $\eta_2= -\frac{2\pi}{L_{1}}p x_{1}$.
For this reason we introduce  $p$ e $l'\rightarrow l$ into the wave function.
Thus under the assumption that the coefficients are limited to a set, as stated in Eq.~\eqref{eq0:18}, the wave function can be expressed in terms of $p$ instead of $H_3$.
Then the wave function, as given by Eq.~\eqref{eq0:16}, becomes,
$$\psi_{n}(x_{1},x_{2},p)=\sum_{l} c_{n,l}\psi_{n,l}(x_{1},x_{2},p)$$ and $$\psi_{n,l}(x_{1},x_{2},p)=A_{n}e^{i\frac{2\pi}{L_{1}}x_{1}}H_{n}\left(\bar{g_{l}}(x_{2})\right)e^{-\frac{1}{2}\left(\bar{g_{l}}(x_{2})\right)^{2}}$$
where  $$\bar{g_{l}}(x_{2})=\sqrt{\frac{2\pi p}{L_{1}L_{2}}}\left( x_{2} + \frac{L_{2}l}{p}\right).$$

It remains to solve Eq.~\eqref{eq0:18} and find the finite number of coefficients, to fully determine the set of $p$ wavefunctions that must be proven orthonormal.
Notice that $p$ is fixed and this defines  $\phi_{n,q}(x_{1},x_{2},p)$, such that $q=0,...,p-1$.
The seek solutions for growing values of $p$ and start with the initial value,  $p=1$.
This is the case treated  by A. Abrikosov in his seminal work where vortices were discovered in superconductivity~\cite{abrikosov56}.
In this case Eq.~\eqref{eq0:18} becomes $c_{n,l+1}=c_{n,l}$, and the solution simply corresponds to all  coefficients equal, regardless of  $l$.
We choose that  $c_{n,l}=c_{n,0}$.
For $p=2$,  Eq.~\eqref{eq0:18} becomes $c_{n,l+2}=c_{n,l}$ and in this case there are two free coefficients, chosen to be
$c_{n,0}$ and $c_{n,1}$, for $l$ even and odd ones, respectively.
Therefore $c_{n,l}=c_{n,0}$ for $l=0,\; \pm 2,\; \pm 4,\; \pm 6,...$ and  $c_{n,l}=c_{n,1}$ for $l=\pm 1,\; \pm 3,\; \pm 4,...$.
For the next case, $p=3$,  Eq.~\eqref{eq0:18} becomes $c_{n,l+3}=c_{n,l}$, and there are tree free coefficients at each Landau level $n$.
Similarly, the choices are $c_{n,0}$, $c_{n,1}$ and $c_{n,2}$.
This gives that $c_{n,l}=c_{n,0}$ for $l=0,\;\pm 3,\; \pm 6,...$,  $c_{n,l}=c_{n,1}$ for $l=\pm 1,\; \pm 4,\; \pm 7,...$ and
$c_{n,l}=c_{n,2}$ for $l=\pm 2,\; \pm 5,\; \pm 8,...$.
The last case explicitly treated here is $p=4$, and in this case, Eq.~\eqref{eq0:18} becomes $c_{n,l+4}=c_{n,l}$.
There are four free coefficients, namely, $c_{n,0}$, $c_{n,1}$, $c_{n,2}$ and $c_{n,3}$.
Hence we introduce the general notation $c_{n,q}$ for the coefficients, such that $q=0,\;1,\;2,...,p-1$, represent the $p$ free and independent coefficients.
Table~\ref{tab1} provides the coefficients for the first four cases, namely, $p=1,\;2, \;3$ and $4$.
For instance, for $p=1$ there is only $q=0$, and consequently, only one free and independent coefficient.
For $p=2$ there are the $q=0$ and $q=1$ cases, and so, two coefficients.
For $p=3$ the three free coefficients are associated to  $q=0,\;1$ and $2$.
Lastly for $p=4$ the four free coefficients are associated to  $q=0,\;1,\;2$ and $3$.
In power of such information we write the most general quasi periodic $\psi_{n}$ function as,
\begin{eqnarray}
\psi_{n,p}(x_{1},x_{2}) = \sum^{p-1}_{q=0}c_{n,q}\phi_{n,q}(x_{1},x_{2},p),
\label{eq0:19}
\end{eqnarray}
where the functions $\phi_{n,q}(x_{1},x_{2},p)$ are defined below,
\begin{equation}
\phi_{n,q}(x_{1},x_{2},p) = A_{n}e^{i\frac{2\pi }{L_1}qx_{1}}\!\! \sum^{+\infty}_{l=-\infty}\!\! e^{i\frac{2\pi }{L_1}plx_{1}} H_{n}(f_{lq})e^{-\frac{1}{2}f_{lq}^{2}},
\label{eq0:20}
\end{equation}
such that $f_{lq}$ is defined by,
\begin{eqnarray}
f_{lq}\equiv\sqrt{\frac{2\pi p}{L_{1}L_{2}}}\left[x_{2} + \frac{L_{2}}{p}(pl+q)\right].
\label{eq0:21}
\end{eqnarray}
%

\begin{table}[t!]
\begin{tabular}{lrrrr}
\hline
\hline
\begin{minipage}[t]{.02\textwidth}\begin{flushleft} \end{flushleft}\end{minipage}&
\begin{minipage}[t]{.02\textwidth}\begin{flushleft} \end{flushleft}\end{minipage}&
\begin{minipage}[t]{.02\textwidth}\begin{flushleft} \end{flushleft}\end{minipage}&
\begin{minipage}[t]{.02\textwidth}\begin{flushleft} \end{flushleft}\end{minipage}&
\begin{minipage}[t]{.02\textwidth}\begin{flushleft}  \end{flushleft}\end{minipage}\\
 $p=1$  \\
\hline
 &$q=0$ \\
&$c_{n,l}=c_{n,0}$	 \\
\hline
\hline
 $p=2$  \\
\hline
&$q=0$	            &$q=1$  \\
&$c_{n,2l}=c_{n,0}$	&$c_{n,2l+1}=c_{n,1}$  \\
\hline
\hline
 $p=3$  \\
\hline
 &$q=0$	            &$q=1$                &$q=2$  \\
 &$c_{n,3l}=c_{n,0}$	&$c_{n,3l+1}=c_{n,1}$ &$c_{n,3l+2}=c_{n,2}$ \\
\hline
\hline
 $p=4$  \\
\hline
 &$q=0$	              &$q=1$                &$q=2$                &$q=3$ \\
 &$c_{n,4l}=c_{n,0}$	  &$c_{n,4l+1}=c_{n,1}$ &$c_{n,4l+2}=c_{n,2}$ &$c_{n,4l+3}=c_{n,3}$ \\
\hline
\hline
\end{tabular}
\label{tab1}
\centering
     \small
     \caption[]{The coefficients $c_{n,q}$, for $p=1,\;2, \;3$ and $4$. }
\end{table}
Next we prove that the $\phi_{n,q}$ functions with the same value of $p$ form an orthonormal set in the rectangular unit cell.
\begin{eqnarray}
\int_{L_1L_2} d^{2}x \; \phi_{m,q'}^{*} \phi_{n,q} = \delta_{qq'}\delta_{nm}.
\label{eq0:22}
\end{eqnarray}
They are degenerate eigenfunctions of the free Hamiltonian with eigenvalues labeled  by $n$, the Landau level index:  $\frac{1}{2m}\vec{D}^{2}\phi_{n,q}=\hbar\omega_{c}(n+\frac{1}{2})\phi_{n,q}$.
For the first Landau level, $n=0$,  $\phi_{0,q}$ satisfies the condition  $D_{+}\phi_{0,q}=0$, where this operator is defined by Eq.~\eqref{eqc:05}.\\

This set of orthonormal functions is central to obtain the results of this paper and for this reason, we have  confirmed that the theoretical prove given below by doing a direct numerical verification.
We have also proven the completeness of the constant $p$ set, as shown below.
Hence we stress that  two functions with distinct values of the trapped flux, say $p_1$ and $p_2$, do not belong to the same set of orthonormal functions, and so, are not orthogonal, as numerically checked.
In fact they belong to two distinct sets of orthonormal functions.
Therefore for a fixed value of $p$, we consider the functions $\phi_{m,q'}$ and $\phi_{n,q}$ and compute the integral  $I_{L_1L_2}$ defined below.
\begin{eqnarray}
&& I_{L_1L_2}=\int_{L_1L_2} d^{2}x \; \phi_{m,q'}^{*} \phi_{n,q} =  \nonumber \\
&& A_{m}A_{n}\sum^{+\infty}_{l'=-\infty}\sum^{+\infty}_{l=-\infty}I_{1}I_{2},
\label{eq0:23}
\end{eqnarray}
where
\begin{eqnarray}
I_{1}&=&\int^{L_{1}}_{0}\; dx_{1} \; e^{i\frac{2\pi}{L_1}[(pl+q)-(pl'+q')]x_{1}},
\label{eq0:24}
\end{eqnarray}
and
\begin{eqnarray}
\!\!\!\! I_{2}&=&\int^{L_{2}}_{0}\!\! dx_{2} \; H_{m}(f'_{l'q'})H_{n}(f_{lq})e^{-\frac{1}{2}\left(f'^{2}_{l'q'}+f_{lq}^{2}\right)},
\label{eq0:25}
\end{eqnarray}
such that $f'_{l'q'}$ is given by Eq.~\eqref{eq0:21} for $l'$ e $q'$.
The integral $I_1$ differs from zero only in case $l=l'$ e $q=q'$.
Then one obtains that $I_{1}=L_{1}\delta_{qq'}$ and this yields that $f'_{l'q'}=f_{lq}$.
A change of variables, $y=f_{lq}$ in Eq.~\eqref{eq0:21} brings a change in the integration variable,  $dx_{2}=\sqrt{\frac{L_{1}L_{2}}{2\pi p}}dy$,
and so, $I_{2}=\sqrt{\frac{L_{1}L_{2}}{2\pi p}}\int^{y_{f}}_{y_{i}} dy H_{m}(y)H_{n}(y)e^{-y^{2}}$,
with the limits of integration given by
$$y_{i}=\sqrt{\frac{2\pi L_{2}}{pL_{1}}}(pl+q) \quad \text{and} \quad y_{f}=\sqrt{\frac{2\pi L_{2}}{pL_{1}}}[p(l+1)+q].$$
Hence the integral of Eq.~\eqref{eq0:23} is expressed as,
\begin{equation}
 I_{L_1L_2}= A_{m}A_{n}\sqrt{\frac{L^{3}_{1}L_{2}}{2\pi p}}\sum^{+\infty}_{l=-\infty}\int^{y_{f}}_{y_{i}} \!\!\!\! dy \; H_{m}(y)H_{n}(y)e^{-y^{2}}.
\label{eq0:26}
\end{equation}
The integral along the $x_2$ direction, which is limited to the unit cell, is extended to the whole axis.
According to the change of variable, $y_{i}$ and $y_{f}$ run from  $l$ until $l+1$, that is, the integrand becomes independent of $l$.
Therefore we write that,
$\sum^{+\infty}_{l=-\infty}\int^{y_{f}}_{y_{i}} dy = \int^{+\infty}_{-\infty} dy$, and in this way,
$\int^{\infty}_{-\infty} dy \; H_{m}(y)H_{n}(y)e^{-y^{2}} = 2^{n}n!\sqrt{\pi}\delta_{nm}$.
Substituting this result into Eq.~\eqref{eq0:26} brings the conclusion that the functions of Eq.~\eqref{eq0:20} are orthogonal. They  become orthonormal according to Eq.~\eqref{eq0:22} by choice of
\begin{equation}
A_{n}=\left(\frac{2\pi p}{L^{3}_{1}L_{2}} \right)^{1/4} \left(2^{n}n!\sqrt{\pi} \right)^{-1/2}.
\label{eq0:27}
\end{equation}
%
\section{Kinetic Energy of free fermions in a magnetic field}
In this section we introduce the second quantization formalism to treat the free fermions in a magnetic field.
The completeness of the set of wave functions under fixed $p$ is verified since it is necessary to prove the anti-commutation relations in real space.
For non-interacting fermions the Hamiltonian is given by,
\begin{equation}
\mathcal{H} = K,
\label{eqa:01}
\end{equation}
where the kinetic energy, $K$, is,
\begin{equation}
K = \int d^{2}x \frac{1}{2m}\left (\vec{D}\Psi\right)^{\dag}\left (\vec{D}\Psi\right).
\label{eqa:02}
\end{equation}
$\Psi$ is a second quantized field and we aim to fill energy levels according to the exclusion principle up to the Fermi surface.
We express the kinetic energy in a more convenient way,
\begin{equation}
K = \frac{1}{2m}\int\!\! d^{2}x \left(\Psi^{\dag} \vec{D}^2\Psi\right) + \frac{\hbar^{2}}{4m}\int\!\! d^{2}x\, \vec{\nabla}^{2}\rho,
\label{eqa:12}
\end{equation}
where $\rho\equiv\Psi^{\dag}\Psi$.
It happens that the second integral vanishes due to the periodicity of the state, and the kinetic energy becomes,
\begin{equation}
K = \frac{1}{2m}\int\!\! d^{2}x \left(\Psi^{\dag} \vec{D}^2\Psi\right).
\label{eqa:12b}
\end{equation}
To derive Eq.~\eqref{eqa:12} we write $\left (\vec{D}\Psi \right )^{\dag} \left (\vec{D}\Psi \right )= \left(i\hbar\vec{\nabla}\Psi^{\dag}-(e/c)\vec{A}\Psi^{\dag}\right)\cdot\vec{\alpha}$, with $\vec{\alpha}\equiv \vec{D}\Psi  $.
Using that $\vec{\nabla}\Psi^{\dag}\cdot\vec{\alpha} = \vec{\nabla}\cdot(\Psi^{\dag}\vec{\alpha})-\Psi^{\dag}(\vec{\nabla}\cdot \vec{\alpha})$,
one gets that  $$ \left (\vec{D}\Psi \right )^{\dag} \left (\vec{D}\Psi \right )= \Psi^{\dag}\left(-i\hbar\vec{\nabla} - (e/c)\vec{A} \right)\cdot\vec{\alpha} + i\hbar\vec{\nabla}\cdot(\Psi^{\dag}\vec{\alpha}).$$
The kinetic energy  becomes,
\begin{equation}
K= \frac{1}{2m}\int d^{2}x \left(\Psi^{\dag} \vec{D}^{2}\Psi\right) + \frac{i\hbar}{2m}\int d^{2}x \, \vec{\nabla}\cdot\left(\Psi^{\dag} \vec{D}\Psi\right),
\label{eqa:13}
\end{equation}
and its complex conjugate,
\begin{equation}
K^{*}= \frac{1}{2m}\int d^{2}x \left(\vec{D}^{2}\Psi\right)^{\dag}\Psi - \frac{i\hbar}{2m}\int d^{2}x \, \vec{\nabla}\cdot\left((\vec{D}\Psi)^{\dag}\Psi\right).
\label{eqa:14}
\end{equation}
The kinetic energy is real thus summing Eqs. \eqref{eqa:13} e \eqref{eqa:14} and dividing by two, gives that,
\begin{eqnarray}
K &=& \frac{1}{4m}\int d^{2}x \Psi^{\dag} \vec{D}^{2}\Psi + \left (\vec{D}^{2}\Psi\right)^{\dag}\Psi \nonumber \\
&+& \frac{i\hbar}{4m}\int d^{2}x \, \vec{\nabla}\cdot\left(\Psi^{\dag} \vec{D}\Psi - \Psi (\vec{D}\Psi)^{\dag}\right).\nonumber
\label{eqa0:14}
\end{eqnarray}
%
\begin{figure}[t!]
\centering
\includegraphics[scale=0.6]{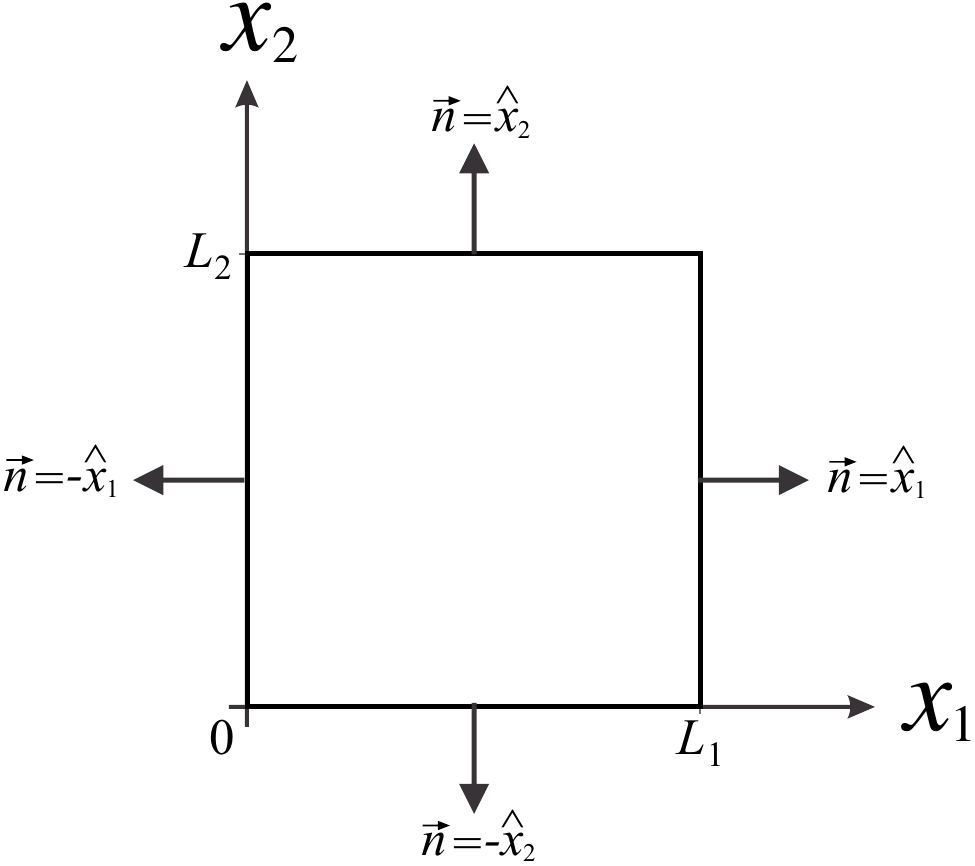}
\caption{The integration scheme and the definition of the vector $\vec{n}$ perpendicular to the surface. }
\label{divergente}
\end{figure}
The two terms in the first integral are shown to be  equal, $\Psi^{\dag} \vec{D}^{2}\Psi=\left(\vec{D}^{2}\Psi\right)^{\dag}\Psi$, and in the second integral it holds that
$\Psi^{\dag} \vec{D}\Psi - (\vec{D}\Psi)^{\dag}\Psi = \vec{\nabla}\left(\Psi^{\dag}\Psi\right)$.
In this way we obtain the symmetrized kinetic energy given in Eq.~\eqref{eqa:12}.


The second integral \eqref{eqa:12} simply vanishes in case of a periodic state, as shown below.
We write it as a surface integral,
\begin{equation}
\frac{\hbar^{2}}{4m}\int\!\! d^{2}x\; \vec{\nabla}^{2}\rho=\frac{\hbar^{2}}{4m} \oint \left(\vec{\nabla}\rho\right)\cdot\vec{n}dl,
\label{eqa:17}
\end{equation}
where $\vec{n}$ is a vector normal to the border lines of the unit cell where the above integration is taken and performed.
using the theorem $\oint_{c} \vec{F}\cdot\vec{n}dl=\int_{S}\, \vec{\nabla}\cdot\vec{F}d^{2}x$ for a vector function $\vec F$ in the plane $(x_{1},x_{2})$.
The integral of Eq.~\eqref{eqa:17} is illustrated in Fig.~\ref{divergente}.
Notice that the line integrals along the path  $dl=dx_{1}$ is obtained by taking  $\vec{n}=-\hat{x}_{2}$ for  $x_{2}=0$, and $\vec{n}=\hat{x}_{2}$ for $x_{2}=L_{2}$.
The integrals along the path  $dl=dx_{2}$ have $\vec{n}=-\hat{x}_{1}$ for  $x_{1}=0$ and $\vec{n}=\hat{x}_{1}$ for $x_{1}=L_{1}$, then Eq.~\eqref{eqa:17} is written as,
\begin{widetext}
\begin{equation}
\frac{\hbar^2}{4m} \oint \left(\vec{\nabla}\rho\right)\cdot\vec{n}dl = \frac{\hbar^2}{4m}\left\{\int^{L_{1}}_{0}dx_{1}\left( \frac{\partial}{\partial x_{2}}\rho(x_{1},L_{2}) - \frac{\partial}{\partial x_{2}}\rho(x_{1},0)\right) + \int^{L_{2}}_{0}dx_{2}\left( \frac{\partial}{\partial x_{1}}\rho(L_{1},x_{2}) - \frac{\partial}{\partial x_{1}}\rho(0,x_{2})\right) \right\}.
\label{eqa:18}
\end{equation}
\end{widetext}
Recall that $\rho$ is periodic, and so, the integrals in opposite sides of the rectangular unit cell annihilate each other, resulting that,
\begin{equation}
\frac{\hbar^{2}}{4m} \oint \; \left(\vec{\nabla}\rho\right)\cdot\vec{n}\;dl = 0.
\label{eqa:19}
\end{equation}

Next we introduce the second quantization formalism to treat the fermionic particles.
Notice that the second quantized field $\Psi_{p}(x_{1},x_{2})$ carries the index $p$ to stress the $p$ fluxons are trapped in the unit cell area, $L_1L_2$.
\begin{equation}
\Psi_{p}(x_{1},x_{2}) \! =\!\! \sum^{\infty}_{n=0} \Psi_{n,p}(x_{1},x_{2}) \!=\!\! \sum^{\infty}_{n=0}\sum^{p-1}_{q=0}c_{n,q}\phi_{n,q}(x_{1},x_{2},p),
\label{eqa:04}
\end{equation}
where the functions  $\phi_{n,q}(x_{1},x_{2},p)$ are given by Eq.~\eqref{eq0:20} and the coefficients $c_{n,q}$ are to be interpreted as destruction operators below.
The second quantized field $\Psi_{p}(x_{1},x_{2})$  satisfies the anti-commutation  relation,
%
\begin{flalign}
&\left\{\Psi_{p}(x_{1},x_{2}),\Psi^{\dag}_{p}(x'_{1},x'_{2}) \right\}=\delta(x_{1}-x'_{1})\delta(x_{2}-x'_{2}), \; \mbox{and}
\label{eqa:05} \\
&\left\{\Psi_{p}(x_{1},x_{2}),\Psi_{p}(x'_{1},x'_{2}) \right\}= 0, \label{eqa:05b}
\end{flalign}
%
assuming that the operators $c_{n,q}$  obey the conditions below.
\begin{flalign}
& \left\{c_{n,q}\,,\,c^{\dag}_{m,q'} \right\}=\delta_{qq'}\delta_{nm},
\label{eqa:06}\\
& \left\{c_{n,q}\,,\,c_{m,q'} \right\}=0. \label{eqa:06b}
\end{flalign}
The completeness relation of the orthonormal set defined by $p$ is behind  Eq.~\eqref{eqa:05} and to prove we start with Eq.~\eqref{eqa:04}.
Using Eq.~\eqref{eqa:06} it follows that,
\begin{widetext}
\begin{eqnarray}
\left\{\Psi_{p}(x_{1},x_{2}),\Psi^{\dag}_{p}\left(x'_{1},x'_{2}\right) \right\} = \sum^{\infty}_{n=0}\sum^{p-1}_{q=0}\phi_{n,q}(x_{1},x_{2},p)\phi^{*}_{n,q}(x'_{1},x'_{2},p),
\label{eqa:08}
\end{eqnarray}
\end{widetext}
and then,
\begin{widetext}
\begin{eqnarray}
\left\{\Psi_{p}(x_{1},x_{2}),\Psi^{\dag}_{p}\left(x'_{1},x'_{2}\right) \right\} = \sqrt{\frac{2\pi p}{L^{3}_{1}L_{2}}} \sum^{p-1}_{q=0} \sum^{+\infty}_{l=-\infty} e^{i\frac{2\pi}{L_1}(pl+q)x_{1}} \sum^{+\infty}_{l'=-\infty} e^{-i\frac{2\pi}{L_1}(pl'+q)x'_{1}} \sum^{\infty}_{n=0}b^{2}_{n}H_{n}(f_{lq})H_{n}(f'_{l'q})e^{-\frac{1}{2}(f_{lq}^{2}+f'^{2}_{l'q})}.
\label{eqa:08b}
\end{eqnarray}
\end{widetext}
The latter is obtained by using $\phi_{n,q}(x_{1},x_{2},p)$, defined by Eqs.~\eqref{eq0:20} and \eqref{eq0:27}, where $b_{n}=\left(2^{n}n!\sqrt{\pi} \right)^{-1/2}$.
Firstly consider the sum in  $n$, which is essentially done over the normalized wavefunctions of the harmonic oscillator, $\psi_{n}(x)=\left(2^{n}n!\sqrt{\pi} \right)^{-1/2}H_{n}(x)e^{\frac{1}{2}x^{2}}$,
where $H_{n}(x)$ are the Hermite polynomials.
Hence the $\psi_{n}(x)$ satisfy $\int^{\infty}_{-\infty}\psi^{*}_{m}(x)\psi_{n}(x)=\delta_{nm}$ and the corresponding completeness relation is $\sum^{\infty}_{n=0}\psi_{n}(x)\psi_{n}(y)=\delta(x-y)$.
Thus the sum over $n$ in Eq.~\eqref{eqa:08} becomes $$\sum^{\infty}_{n=0}b^{2}_{n}H_{n}(f_{lq})H_{n}(f'_{l'q})e^{-\frac{1}{2}(f_{lq}^{2}+f'^{2}_{l'q})}=\delta(f_{lq}-f'_{l'q}),$$ and using  Eq.~\eqref{eq0:21}, one
obtains that $\sum^{\infty}_{n=0}b^{2}_{n}H_{n}(f_{lq})H_{n}(f'_{l'q})e^{-\frac{1}{2}(f_{lq}^{2}+f'^{2}_{l'q})}=\delta\left(\sqrt{\frac{2\pi p}{L_{1}L_{2}}}(x_{2}-x'_{2}) + L_{2}(l-l') \right)$. Considering that  $x_{2}$ e $l$ are independent variables it holds that the delta function is centered at  $x_2=x^{\prime}_2$ and $l=l'$.
Therefore  we have the following completeness relation.
\begin{equation}
\sum^{\infty}_{n=0}\!b^{2}_{n}H_{n}(f_{lq})H_{n}(f'_{l'q})e^{-\frac{1}{2}(f_{lq}^{2}+f'^{2}_{l'q})}=\sqrt{\frac{L_{1}L_{2}}{2\pi p}}\delta\!\left(x_{2}-x'_{2}\right),
\label{eqa:09}
\end{equation}
where we have used the property $\delta(a(x_2-x'_2))=\frac{1}{a}\delta(x_2-x'_2)$, $a=\sqrt{\frac{2\pi p}{L_{1}L_{2}}}$.
Using Eq.~\eqref{eqa:09}, the  Eq.~\eqref{eqa:08} becomes,
\begin{widetext}
\begin{equation}
\left\{\Psi_{p}(x_{1},x_{2}),\Psi^{\dag}_{p}\left(x'_{1},x'_{2}\right) \right\} = \frac{1}{L_{1}} \sum^{p-1}_{q=0} \sum^{+\infty}_{l=-\infty} e^{i\frac{2\pi}{L_1}(pl+q)(x_{1}-x'_{1})}\delta\!\left(x_{2}-x'_{2}\right)
\label{eqa:10}.
\end{equation}
\end{widetext}
 Next the sum $S_{p}=\sum^{p-1}_{q=0} \sum^{+\infty}_{l=-\infty} e^{i\frac{2\pi}{L_1}(pl+q)(x_{1}-x'_{1})}$ must be analysed. For $p=1$ the sum becomes $S_{1}= \sum^{+\infty}_{l=-\infty} e^{i\frac{2\pi}{L_1}l(x_{1}-x'_{1})}$, for $p=2$, it is  $S_{2}=\sum^{+\infty}_{l=-\infty} \left(e^{i\frac{2\pi}{L_1}(2l)(x_{1}-x'_{1})}+e^{i\frac{2\pi}{L_1}(2l+1)(x_{1}-x'_{1})}\right)$, and for  $p=3$ it is given by $S_{3}=\sum^{+\infty}_{l=-\infty} \left(e^{i\frac{2\pi}{L_1}(3l)(x_{1}-x'_{1})}+e^{i\frac{2\pi}{L_1}(3l+1)(x_{1}-x'_{1})}\right)+ \sum^{+\infty}_{l=-\infty}e^{i\frac{2\pi}{L_1}(3l+2)(x_{1}-x'_{1})}$.
 Notice that the total sum runs over all integers, and in this way, one can write that  $$S_{p}=\sum^{p-1}_{q=0} \sum^{+\infty}_{l=-\infty} e^{i\frac{2\pi}{L_1}(pl+q)(x_{1}-x'_{1})}=\sum^{+\infty}_{r=-\infty} e^{i\frac{2\pi}{L_1}r(x_{1}-x'_{1})}.$$ Hence we have reached the second completeness relation,
\begin{equation}
S_{p}=\sum^{+\infty}_{r=-\infty} e^{i\frac{2\pi}{L_1}r(x_{1}-x'_{1})}=L_{1}\delta\!\left(x_{1}-x'_{1}\right),
\label{eqa:11}
\end{equation}
that once added to  Eq.~\eqref{eqa:10}, gives the anti-commuting relation given by Eq.~\eqref{eqa:05}, thus proving its correctness.
\subsection{The energy and the number of particles in terms of the operators $c_{n,q}$}
The kinetic energy $K$, given by  Eq.~\eqref{eqa:12b}, and $\Psi=\Psi_{p}$ given by Eq.~\eqref{eqa:04}, once put together render the kinetic energy expressed in terms of the operators $c_{n,q}$.
The following identity
\begin{equation}
\frac{1}{2m}\int\!\! d^{2}x \; \left(\Psi^{\dag} \vec{D}^2\Psi\right) \!=\! \sum_{n,m}\sum_{q,q'}E_{n}c^{\dag}_{m,q'}c_{n,q}\int\!\! d^{2}x\phi^{*}_{m,q'}\phi_{n,q},
\label{eqa:15}
\end{equation}
added to the orthonormality condition, Eq.~\eqref{eq0:22}, restricts the sums to $m=n$ e $q=q'$.
\begin{equation}
\frac{1}{2m}\int\!\! d^{2}x \; \left(\Psi^{\dag} \vec{D}^2\Psi\right) \!=\! \sum^{\infty}_{n=0}\sum^{p-1}_{q=0}E_{n}c^{\dag}_{n,q}c_{n,q},
\label{eqa:16}
\end{equation}
where $E_{n}$ is given by Eq.~\eqref{eq0:05}. Hence the Hamiltonian of Eq.~\eqref{eqa:01} acquires the diagonal form given by,
\begin{equation}
\mathcal{H} = \sum^{\infty}_{n=0}\sum^{p-1}_{q=0}E_{n}c^{\dag}_{n,q}c_{n,q}.
\label{eqa:20}
\end{equation}
The operator that determines the number of particles is also diagonal and given by,
\begin{equation}
\mathcal{N} = \int\!\! d^{2}x \Psi^{\dag} \Psi=\sum^{\infty}_{n=0}\sum^{p-1}_{q=0} c^{\dag}_{n,q}c_{n,q}.
\label{eqa:12c}
\end{equation}


\begin{figure*}[t!]
\centering
\includegraphics[scale=0.6]{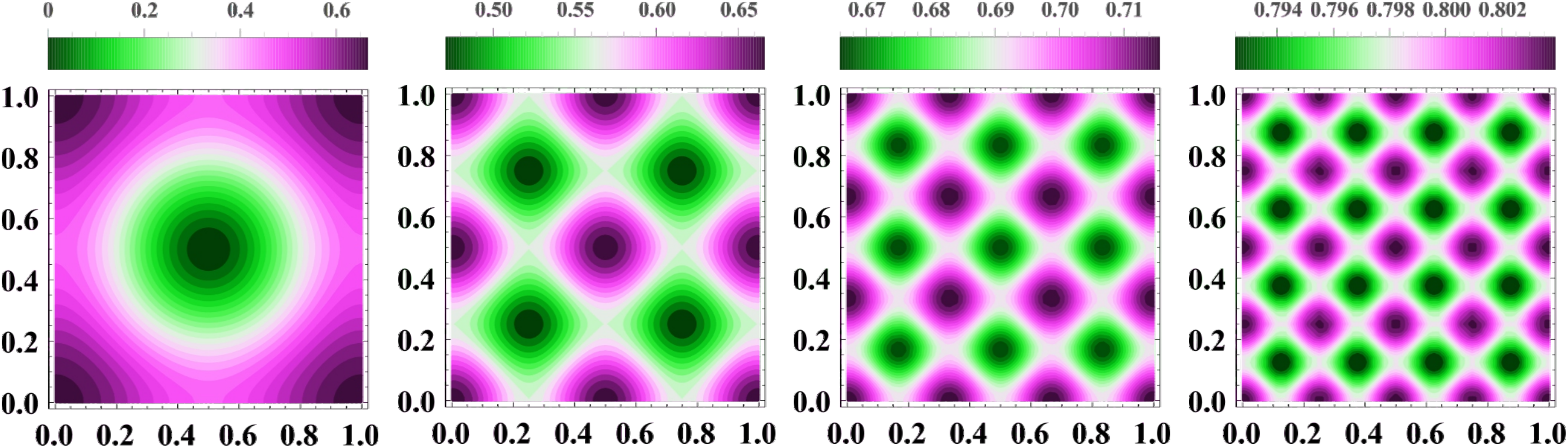}
\caption{The spatial electronic density in the unit cell ($x_{1}/L_{1}$, $x_{2}/L_{2}$) for the case of the completely filled  first Landau levels ($n=0$, sum over $q$). The number of trapped magnetic fluxons is equal to the number of particles, $p$, and the plots are shown from left to right according to $p=1,2,3$ and $4$.}
\label{fig1}
\end{figure*}

\begin{figure*}[t!]
\centering
\includegraphics[scale=0.6]{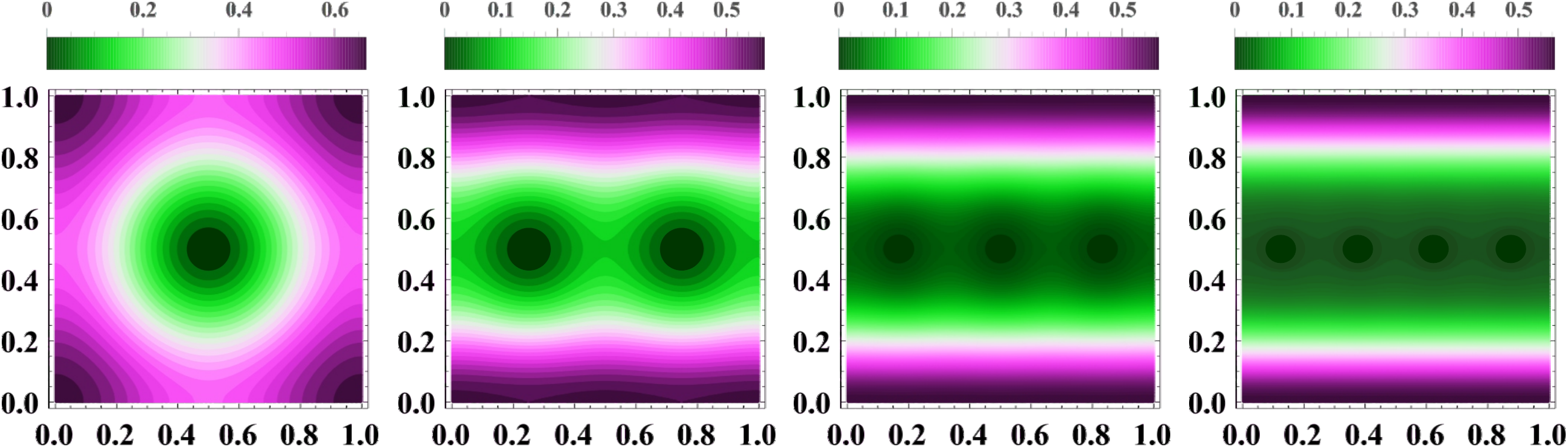}
\caption{The spatial electronic density is shown in the unit cell ($x_{1}/L_{1}$, $x_{2}/L_{2}$) for the case of the partially  filled  first Landau levels ($n=0$, $q=0$).
The number of trapped magnetic fluxons is $p$ whereas there is only one particle in the unit cell.
The plots are shown from left to right according to $p=1,2,3$ and $4$.}
\label{fig2}
\end{figure*}

\begin{figure*}[t!]
\centering
\includegraphics[scale=0.52]{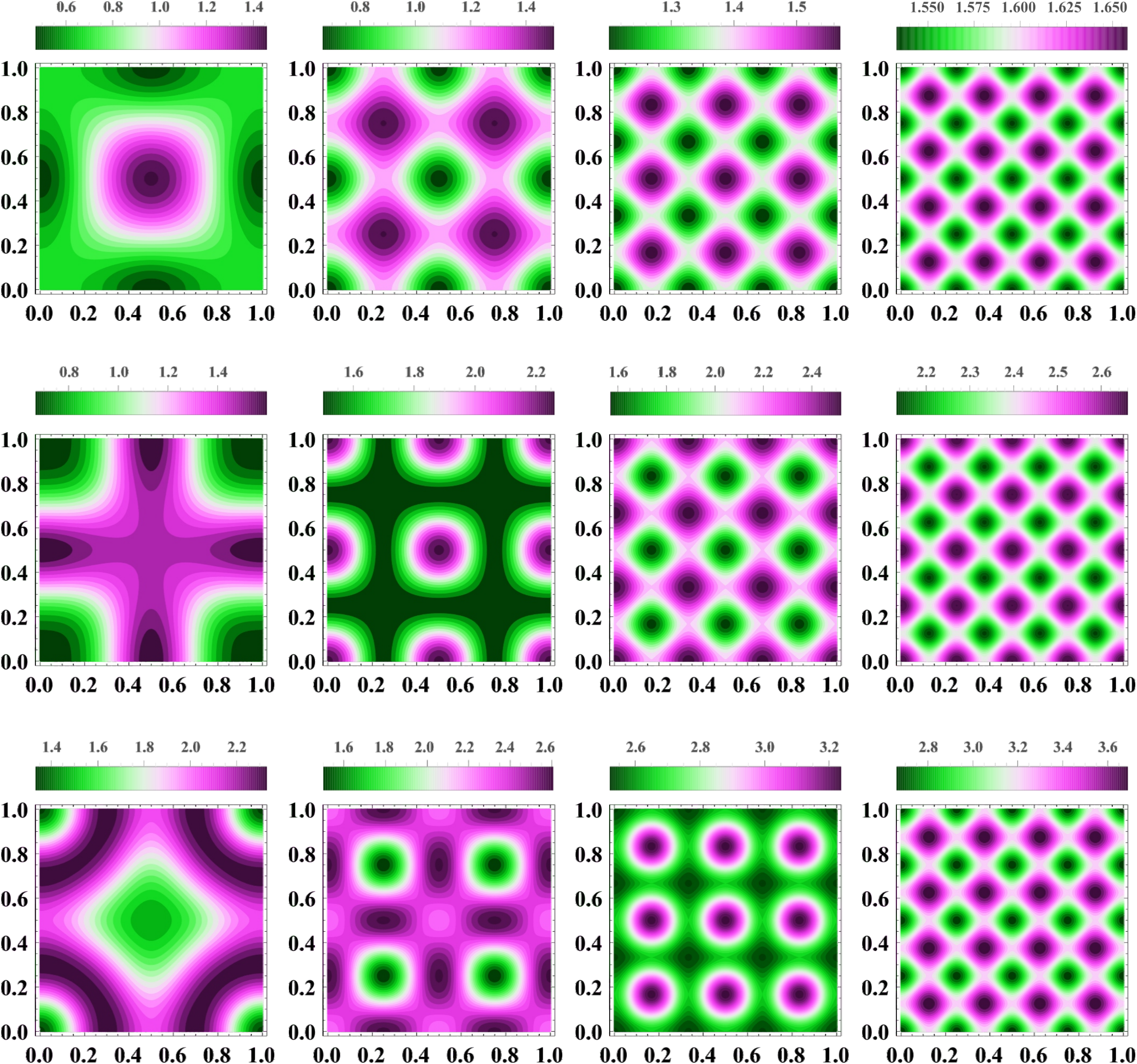}
\caption{ The spatial electronic density $\left\langle \Phi |\rho_{p}|\Phi\right\rangle$, obtained from Eq.~\eqref{D05}, is shown in the unit cell $x_{1}/L_{1}$, ($x_{2}/L_{2}$) for the case of completely filled  Landau levels.
The rows correspond to the sum over Landau levels, $n'=0,1$, $n'=0,1,2$ and $n'=0,1,2,3$ from top to bottom,  and
are associated to the maximum Landau level $n=1,2$ and $3$, respectively.
The columns run from left to right the number of particles in the unit cell, $p=1,2,3$ and $4$, which is the same as the number of trapped flux.
Then the plots are labeled by ($p,n$) where $p$ and $n$ are integers and the total number of particles associated to each one is $N(p,n)=(n+1)p$.
Interestingly the density shows $p^2$ maxima (or minima) independent of $n$, the number of filled Landau levels.
Hence ratio between the number of particles and the number of density maxima (minima) is given by $(n+1)/p$.
The ratio of each of the  plots displayed are given in Table~\ref{tab-ratio}.}
\label{fig3}
\end{figure*}

\subsection{The wave function and the spatial electronic density}
The fermions fill all the levels up to the highest one, and so, the first $n'$ Landau levels, $n=0,1,2,...,n'-1$, each with $p$ electrons, and the last Landau level  $n'$ can be partially filled with $p'$ electrons, hence $p'< p$.
Therefore the wave function $|\Phi\rangle$ representing this state has the form $|\Phi\rangle  = \prod_{n,q}c^{\dag}_{n,q}|0\rangle$
where $c^{\dag}_{n,q}$ is a creation operator, the index  $n\in[0, n']$ represents the Landau levels, and the index $q$, $0\leq q \leq \tilde{p}-1$, runs over
$\tilde{p}$, which is the number of particles in each Landau level $n$.
An explicit version of the wave function is given below.
\begin{widetext}
\begin{equation}
|\Phi\rangle  = \underbrace{c^{\dag}_{n',p'-1}\cdots c^{\dag}_{n',1}c^{\dag}_{n',0}}_{(\text{level $n=n'$ incomplete})} \overbrace{c^{\dag}_{n'-1,p-1}\cdots c^{\dag}_{n'-1,1}c^{\dag}_{n'-1,0}}^{(\text{level $n=n'-1$ complete})} \cdots \underbrace{c^{\dag}_{1,p-1}\cdots c^{\dag}_{1,1}c^{\dag}_{1,0}}_{(\text{level $n=1$ complete})}  \overbrace{c^{\dag}_{0,p-1}\cdots c^{\dag}_{0,1}c^{\dag}_{0,0}}^{(\text{level $n=0$ complete})}|0\rangle.
\label{D03}
\end{equation}
\end{widetext}
The state $|\Phi\rangle$ is automatically normalized, $\langle\Phi|\Phi\rangle = 1$ and has a fixed number of particles, Eq.~\eqref{eqa:12c}, since,
\begin{equation}
\mathcal{N}|\Phi\rangle = N |\Phi\rangle,\, \mbox{where} \, N = n^{\prime}p + p^{\prime}
\label{eqa:12d}
\end{equation}

The spatially distributed electronic density is given by the expectation value $\left\langle \Phi |\rho_{p}|\Phi\right\rangle$, where  $\rho_{p}\equiv\Psi^{\dag}_{p}\Psi_{p}$ and $\Psi_{p}$
is the operator associated to the periodic  density, that acquires the following form.
\begin{equation}
\rho_{p}  = \sum^{\infty}_{m=0}\sum^{\infty}_{n=0}\sum^{p-1}_{q'=0}\sum^{p-1}_{q=0}c^{\dag}_{m,q'}c_{n,q}\phi^{*}_{m,q'}\phi_{n,q}.
\label{D04}
\end{equation}
Then the expectation value is obtained used the anti-commutation condition of Eq.~\eqref{eqa:06}.
\begin{equation}
\left\langle \Phi |\rho_{p}|\Phi\right\rangle  = \sum^{n^{\prime}-1}_{n=0}\sum^{p-1}_{q=0}|\phi_{n,q}(p)|^{2} + \sum^{p^{\prime}-1}_{q=0}|\phi_{n^{\prime},q}(p)|^{2}.
\label{D05}
\end{equation}
This calculation is straightforward once observed that  there is no contribution to the expectation value in case that the $\rho_p$  operators,  $c^{\dag}_{m,q'}$ or $c_{n,q}$,  are not contained in  $|\Phi\rangle$.
Notice that the  functions belonging to the orthonormal set are being explicitly expressed with its $p$ dependence, namely as $\phi_{n,q}(p)$, although this notation is implicit elsewhere.\\

Fig.~\ref{fig1} shows the spatial electronic density of the completely filled  first Landau levels ($n=0$).
In this case the wavefunction \eqref{D03} is $|\Phi\rangle  = c^{\dag}_{0,p-1}\cdots c^{\dag}_{0,1}c^{\dag}_{0,0}|0\rangle$, and so the spatial density becomes $\left\langle \Phi |\rho_{p}|\Phi\right\rangle =\sum^{p-1}_{q=0}|\phi_{0,q}(p)|^{2}$, obtained from Eq.~\eqref{D05}.
The number of trapped flux and the number of particles is the same and equal to $p$.
Interestingly the density displays an egg-box pattern with $p^2$ maxima (minima).
Hence each maxima (minima) can be associated to a fraction $1/p$ of a  particle.
Notice that as $p$ increases, the difference between the maximum and the minimimum density shrinks.
This is a consequence of the sum over the $|\phi_{0,q}(p)|^{2}$ wave functions which adds more positive contributions as $p$ increases.
for this reason the first plot, which corresponds to $p=1$, is able to  reach zero density, as the sum over $q$ is absent.\\

Fig.~\ref{fig2} shows the spatial electronic density for the partially filled  first Landau levels ($n=0$) such that only the $q=0$ state is present.
In this case the wave function \eqref{D03} contains a single particle, $|\Phi\rangle  = c^{\dag}_{0,0}|0\rangle$, and the spatial density becomes $\left\langle \Phi |\rho_{p}|\Phi\right\rangle =|\phi_{0,0}(p)|^{2}$, obtained from Eq.~\eqref{D05}.
The number of trapped flux and the number of particles is not the same, the former is $p$ whereas the latter is one.
The density displays  $p$ zeros, which is the number of trapped flux in the unit cell.
Thus each zero in the density is associated to $1/p$ particles.
The first plot of Fig.~\ref{fig1} and Fig.~\ref{fig2} coincide as both correspond to $p=1$  and $q=0$.\\

Fig.~\ref{fig3} shows the remarkable fact that the electronic density of  $N=(n+1)p$ particles displays an egg-box pattern with $p^{2}$ maxima (minima).
There are $p$  particles (and  fluxons) in each Landau level and $n$ filled Landau levels, thus making a total of $N=(n+1)p$ particles (fluxons).
From top to bottom the  sum is over the Landau levels $n'=0,1$, $n'=0,1,2$ and $n'=0,1,2,3$, thus defining
the maximum Landau level $n=1,2$ and $3$.
Hence the unit cell is shown in units of ($x_{1}/L_{1}, x_{2}/L_{2}$), such that the last level of Eq.~\eqref{D05}, $n^{\prime}=n$, is also filled,  $p^{\prime}=p$.
The columns run from left to right and  the number of particles in the unit cell ranges $p=1,2,3$ and $4$.
Notice that the scales of the plots do not coincide although the same set of colors is used.
The case where only the lowest Landau level is considered, $n=n'=0$, is treated in Fig.~\ref{fig1}.
Therefore the first, second and  third rows correspond to the densities, $\left\langle \Phi |\rho_{p}|\Phi\right\rangle  = |\phi_{0,0}(p)|^{2}+|\phi_{1,0}(p)|^{2}$, $\left\langle \Phi |\rho_{p}|\Phi\right\rangle  = |\phi_{0,0}(p)|^{2}+|\phi_{1,0}(p)|^{2}+|\phi_{2,0}(p)|^{2} $, and $\left\langle \Phi |\rho_{p}|\Phi\right\rangle  = |\phi_{0,0}(p)|^{2}+|\phi_{1,0}(p)|^{2}+|\phi_{2,0}(p)|^{2} +|\phi_{3,0}(p)|^{2} $, respectively.
Remarkably, the ratio between the total number of particles and the number of maxima (minima) is fractional, and given by $(n+1)/p$, as shown in Table~\ref{tab-ratio}.

\begin{table}[h!]
\begin{tabular}{|l|r|r|r|r|}
\hline
\hline
\begin{minipage}[t]{.05\textwidth}\begin{flushleft} $ $\end{flushleft}\end{minipage}&
\begin{minipage}[t]{.07\textwidth}\begin{flushright} $p=1$\end{flushright}\end{minipage}&
\begin{minipage}[t]{.07\textwidth}\begin{flushright} $p=2$\end{flushright}\end{minipage}&
\begin{minipage}[t]{.07\textwidth}\begin{flushright} $p=3$\end{flushright}\end{minipage}&
\begin{minipage}[t]{.07\textwidth}\begin{flushright} $p=4$ \end{flushright}\end{minipage}\\
\hline
$n=0$ &$1$	&$1/2$ &$1/3$ &$1/4$ \\
\hline
$n=1$ &$2$	&$1$ &$2/3$ &$1/2$ \\
\hline
$n=2$ &$3$	&$3/2$ &$1$ &$3/4$ \\
\hline
$n=3$ &$4$	&$2$ &$4/3$ &$1$ \\
\hline
\hline
\end{tabular}
\centering
     \small
     \caption[]{The ($n,p$) above elements are associated to the Landau level $n$, and the number of particles in each Landau level, $p$, Indices run  $n=0,1,2,3$ and $p=1,2,3,4$, respectively.
      Notice that all levels are filled up to $n$ which gives for the total number of particles, $p(n+1)$.  The number of density maxima (minima) observed in Figs.~\ref{fig1} and ~\ref{fig3} is $p^2$.
      Therefore the ratio between the total number of particles and the observed maxima (minima) is fractional and given by $(n+1)/p$.}
\label{tab-ratio}
\end{table}

\subsection{de Haas-van Alphen oscillations}

In this section we show that the present formalism describes the well-known results of de Hass-van Alphen oscillations, which essentially describe how the energy and the magnetization change according to the magnetic field.
There are $N$ spinless fermions distributed in $n'+1$ Landau levels, the first $n'$ ones, $n=0,1,2,...,n'-1$, are totally filled and the highest one, $n'$, can be partially filled with $p^{\prime}$ electrons.
Recall that in the last sections we have developed a set of orthonormal wave functions $\phi_{n,q}(x_{1},x_{2},p)$, defined by Eqs.~\eqref{eq0:20} and  \eqref{eq0:21}, where $p$ is the number of trapped magnetic fluxons in the unit cell.
The  periodicity of $\phi_{n,q}(x_{1},x_{2},p)$ has required  that  $p=H_3 L_1 L_2/\Phi_0$, according to Eq.~\eqref{eq0:14}.
The second quantization formalism shows that there are $p$ available states for any Landau level since $q=0,\dots, p-1$ but not all of them needs to be filled.
Therefore the number of fermions remains free to be determined.
for this a new parameter $\lambda$ is introduced  and is  associated to the highest Landau level, $n'$, that is only partially filled.
\begin{eqnarray}\label{lambda}
p'= \lambda p, \; \mbox{where} \; \lambda \in [0,1)
\end{eqnarray}
%
\begin{figure*}[t!]
\centering
\includegraphics[scale=0.56]{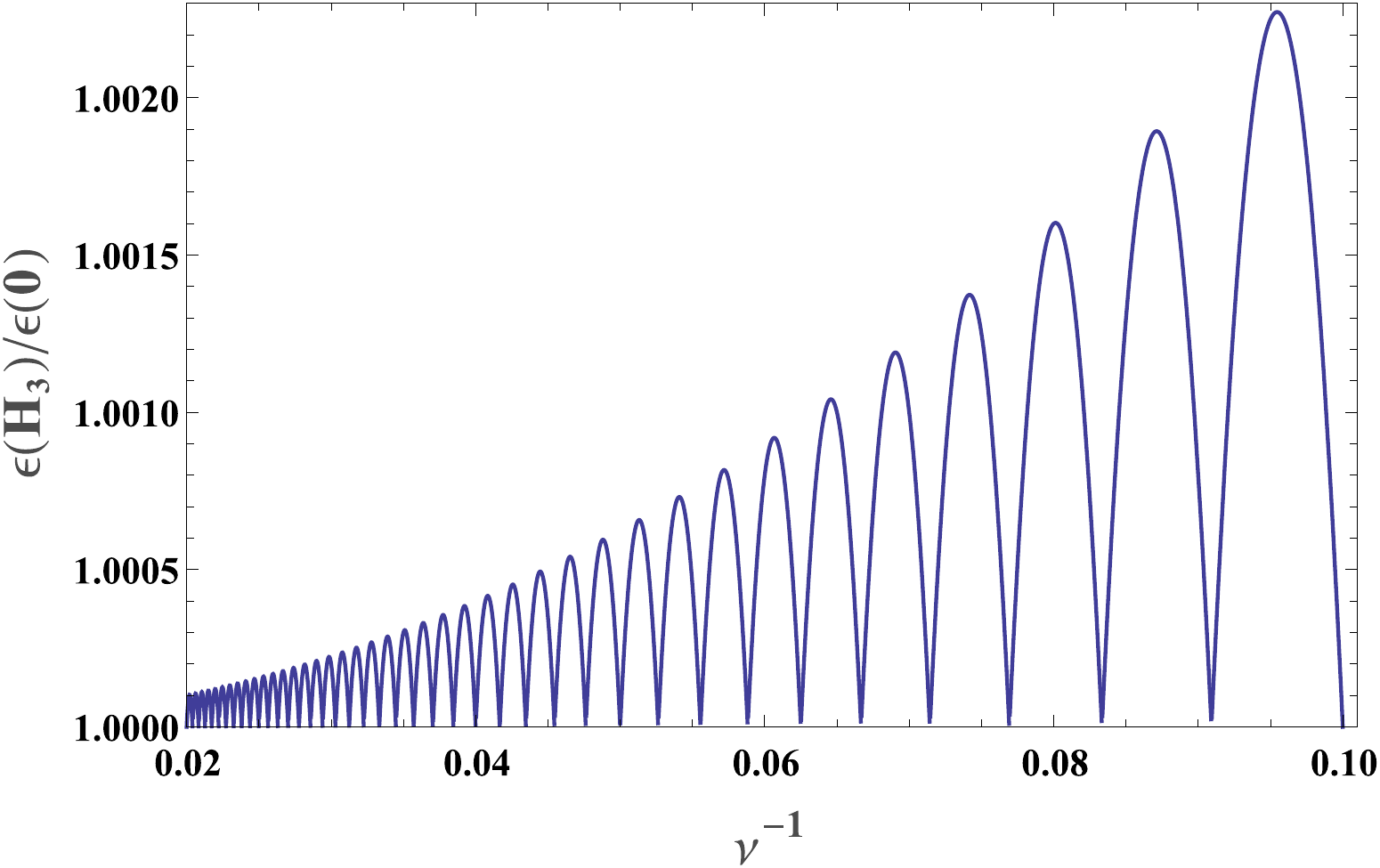}
\includegraphics[scale=0.56]{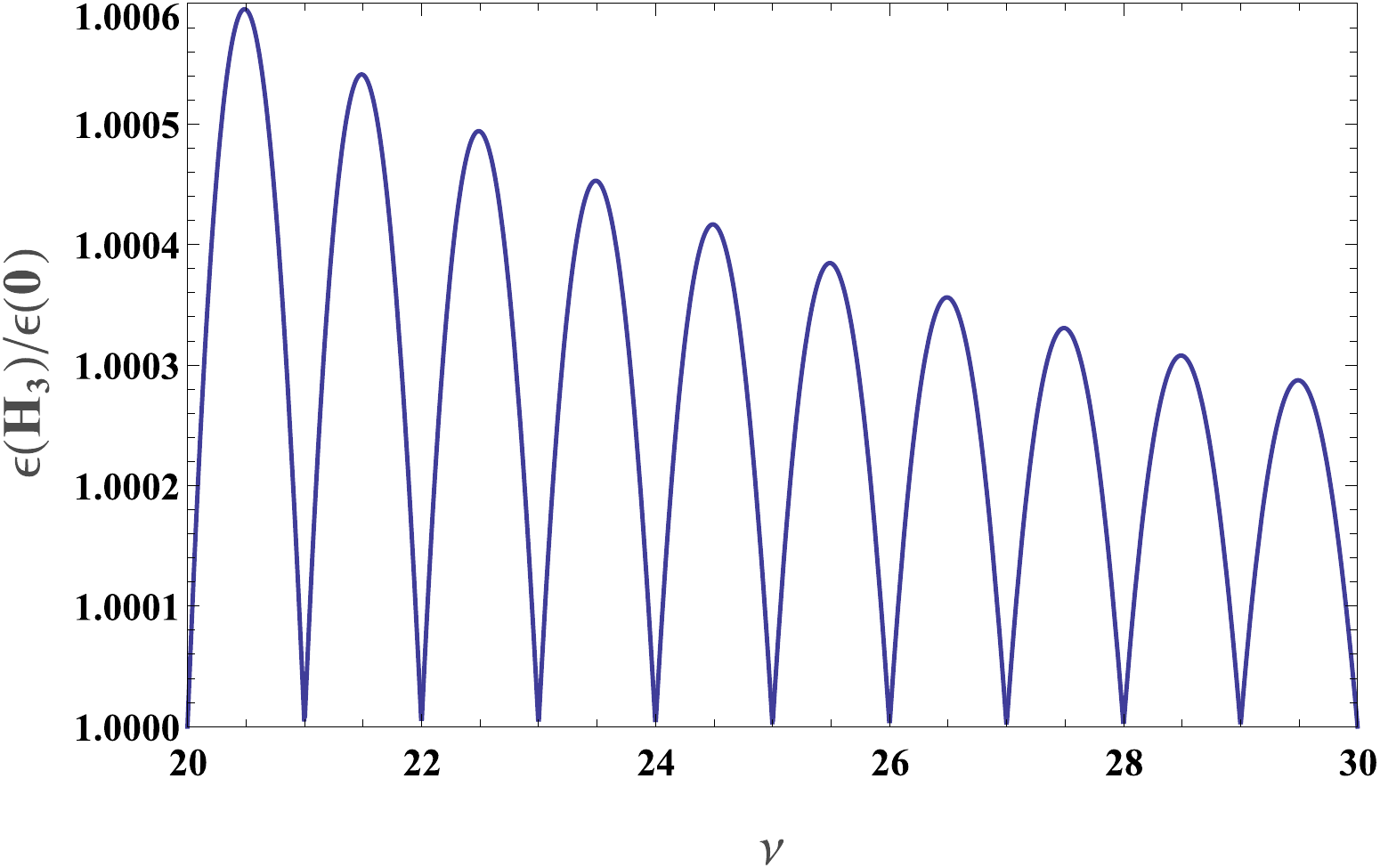}
\caption{The oscillations of the energy, normalized by its zero magnetic field, is shown here, as obtained from  Eq.~\eqref{eqE:16}.
The left plot shows this energy ratio as a function of the inverse of the filling factor, $\nu^{-1} = H_{3}/H'_{3}$, whereas the right one is with respect to the filling factor,  $\nu = H'_{3}/H_{3}$.
$\nu$ describes the number of filled Landau levels and $ H'_{3}$ is the required to locate all electrons in the lowest Landau level.
Notice the periodicity in both $H_{3}$ and $1/H_{3}$.}
\label{fig4}
\end{figure*}
We define a filling variable,
\begin{equation}
\nu \equiv \frac{N}{p},
\label{eqE:06}
\end{equation}
that determines the number of occupied Landau levels according to the number of particles.
\begin{equation}
\nu = n' + \lambda.
\label{eqE:07}
\end{equation}
Notice that $\nu$ and $\lambda$ are continuous variables whereas $n'$ is a discrete variable.
A new  critical field $H'_{3}$ ia defined associated to the density of particles, $N/A$,
\begin{equation}
H'_{3}\equiv \frac{N}{A}\Phi_{0},
\label{eqE:09}
\end{equation}
where $A=L_{1}L_{2}$ is the unit cell area.
As the applied field  $H_{3}$ is associated to $p$ according to  Eq.~\eqref{eq0:14}, and Eq.~\eqref{eqE:06} holds, one obtains that,
\begin{equation}
\nu =\frac{H'_{3}}{H_{3}}.
\label{eqE:08}
\end{equation}
Combining  Eqs.~\eqref{eq0:04}, \eqref{eq0:15}, \eqref{eqE:08} and  \eqref{eqE:09}, one obtains a new way to write the filling factor, $\nu$.
\begin{equation}
\nu =\frac{2\pi N\hbar^{2}}{Am}\frac{1}{\hbar\omega_{c}}.
\label{eqE:10}
\end{equation}
The Landau levels are filled in multiples of  $H'_{3}$ since  $n'=\left[\frac{H'_{3}}{H_{3}}\right]$, and  the incomplete filling $\lambda$ becomes,
\begin{equation}
\lambda=\frac{H'_{3}}{H_{3}}-\left[\frac{H'_{3}}{H_{3}}\right].
\label{eqE:11}
\end{equation}
The notation $[\alpha]$ means to take the largest equal or smaller integer contained in a number $\alpha$.\\

Further  physical insight about the filling factor $\nu$ is achieved by introducing the magnetic length $l_0=\sqrt{\hbar c /e H_3}$.
This length describes the semi-classical orbit of the particle in presence of the magnetic field.
It stems from the centripetal force, $mv^2/r=qvH_3/c$, where $v$ is the velocity and $r$ the radius of the circular orbit.
Added to Bohr's angular momentum condition, $m v r =n_0\hbar$, where $n_0$ is an integer, it leads that the orbits are quantized, $r=\sqrt{n_0}l_0$.
Two possible areas are associated to the particle, namely, the disk associated to the magnetic  orbit, $\pi l_0^2$, and the area defined by the electronic density, $A/N$.
The filling factor corresponds to their ratio, $\nu = 2\pi l_0^2/(A/N)$.
Previously we found that the filling factor is large, $\nu \gg 1$, in case that many Landau levels are filled, $n' \gg 1$.
Therefore in this case there are many overlapping  orbits and this semi-classical picture no longer holds and should be abandoned.
This shows that the present lattice state  is beyond the semi-classical view.\\

The total energy of the fermionic particles requires the summation over all levels and this is readily obtained  from the Hamiltonian of Eq.~\eqref{eqa:20} by computing the energy expectation value,
\begin{eqnarray}
\hspace{-0.15cm}  E_{p} &=& \langle\Phi|\mathcal{H}|\Phi\rangle = \left(E_{0}p + E_{1}p +\ldots +E_{n'-1}p\right) +E_{n'}p' \nonumber \\
       &=& p\sum^{n'-1}_{n=0}E_{n} + E_{n'}p',
\label{eqE:02}
\end{eqnarray}
where $E_n=\hbar\omega_{c}\left(n+1/2\right)$ is given by Eq.~\eqref{eq0:05}, and $E_{n'}=\hbar\omega_{c}\left(n'+1/2\right)$.
In summary, total energy $E_{p} $ takes the contribution of $n'$ fully filled Landau levels, $p\sum^{n'-1}_{n=0}E_{n}$, added to the highest level, $n'$, that may be partially filled,  $E_{n'}p'$.
Introducing Eq.~\eqref{lambda} and the sum $\sum^{n'-1}_{n=0}\left(n+1/2\right)=n'^{2}/2$, one obtains that,
\begin{equation}
E_{p} = \left[\frac{n'^{2}}{2} + \left(n'+\frac{1}{2}\right)\lambda \right]\hbar\omega_{c}p.
\label{eqE:05}
\end{equation}
The energy $E_{p}$ is shown to be explicitly periodic with respect to the filling factor $\nu$, or to its inverse $\nu^{-1}$, the latter being proportional to the applied field $H_3$.
We combine the Eqs.~\eqref{eqE:05} and \eqref{eqE:06} to obtain the energy per particle  as a function of the applied field.
\begin{equation}
\epsilon(H_{3}) \equiv \frac{E_{p}}{N} = \left(\frac{n'^{2}}{2} + \left(n'+\frac{1}{2}\right)\lambda \right)\frac{\hbar\omega_{c}}{\nu}.
\label{eqE:12}
\end{equation}
Using Eq.~\eqref{eqE:10} one obtains that,
\begin{equation}
\epsilon(H_{3}) = \epsilon(0) + \frac{\hbar\omega_{c}}{2\nu}\lambda(1-\lambda),
\label{eqE:14}
\end{equation}
where $\epsilon(0)$ is the energy per particle in the absence of field,
\begin{equation}
\epsilon(0) = \frac{N\pi \hbar^{2}}{Am}.
\label{eqE:15}
\end{equation}
This is done by firstly casting the energy as $\epsilon(H_{3}) = \left(n'^{2}+2n'\lambda + \lambda \right)\frac{\hbar\omega_{c}}{2\nu}$, and then,
adding a term $\pm\lambda^{2}$ inside the parenthesis to obtain that  $\epsilon(H_{3}) = \left((n'+\lambda)^{2} + \lambda -\lambda^{2} \right)\frac{\hbar\omega_{c}}{2\nu}$.
Next using that $\nu = n' + \lambda$, Eq.~\eqref{eqE:07}, one obtains that,
\begin{equation}
\epsilon(H_{3}) = \frac{1}{2}\nu\hbar\omega_{c} + \frac{\hbar\omega_{c}}{2\nu}\lambda(1-\lambda).
\label{eqE:13}
\end{equation}

We briefly review the derivation of the  average energy per particle without the applied field, given by  Eq.~\eqref{eqE:15}.
This can be achieved by direct arguments without the inclusion of the applied field.
By integrating over the Fermi surface disk, one obtains that $N/A=k_F^2/4\pi$ and that the total energy per particle is $E_T/N=E_F/2$, $E_F=(\hbar k_F)^2/2m$.
Hence one finds the zero field energy per particle of Eq.~\eqref{eqE:15}, that is, $E_T/N=\epsilon(0)$.
Hence the ratio between the energy in presence of a field, $\epsilon(H_{3})$, and without a field,  $\epsilon(0)$, becomes,
\begin{equation}
\frac{\epsilon(H_{3})}{\epsilon(0)} = 1 + \left(\frac{H_{3}}{H'_{3}}\right)^{2}\lambda(1-\lambda).
\label{eqE:16}
\end{equation}
The above expression is plotted in Fig.~\ref{fig4} and shows the oscillations with respect to the filling factor.
Since $\nu^{-1}= H_3/H^{\prime}_3$ the left plot shows that the period of oscillations increases for increasing  $H_{3}$.
The  amplitude of oscillations smoothly increases proportional to  $H_{3}^{2}$ according to Eq.~\eqref{eqE:16}, while $\lambda$ is a periodic function of $1/H_{3}$.
The same data is also plotted in the right plot,  in this case with respect to $\nu$, and both plots show a periodicity.

\begin{figure}[t!]
\centering
\includegraphics[scale=0.58]{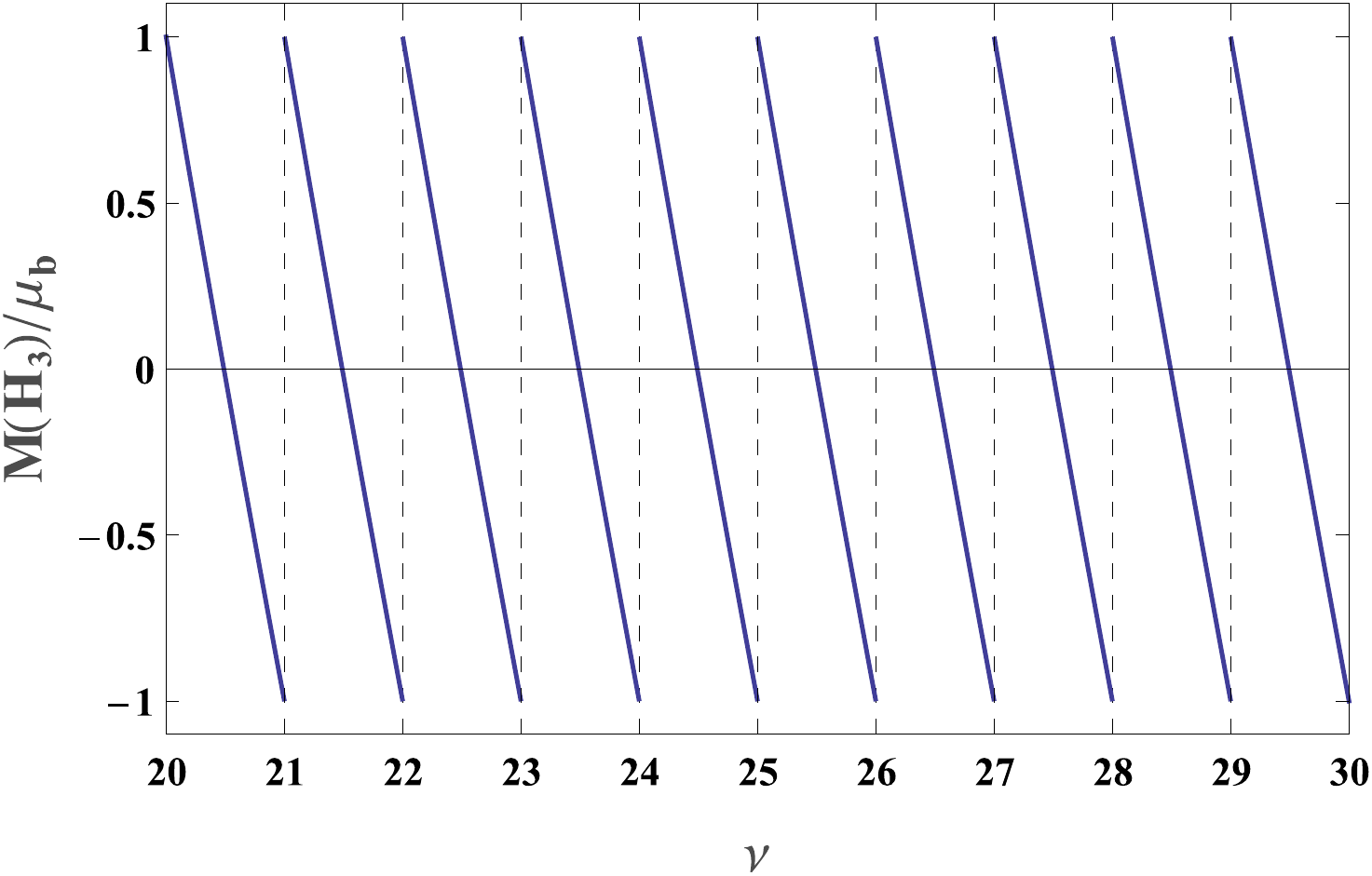}
\caption{The magnetization, given by Eq.~\eqref{eqE:20}, as a function of $\nu = H'_{3}/H_{3}$, where $H'_{3} = N/(A\Phi_{0})$, according to  Eq.~\eqref{eqE:09}.
Notice the periodicity of the magnetization with respect to $1/H_{3}$.}
\label{fig5}
\end{figure}

The magnetization $M$ is obtained from,
\begin{equation}
M = -\frac{\partial \epsilon(H_{3})}{\partial H_{3}},
\label{eqE:17}
\end{equation}
where $\epsilon(H_{3})$ is given by Eq.~\eqref{eqE:14}, that once combined with  Eqs.~\eqref{eq0:04} and \eqref{eqE:11}, gives that,
\begin{widetext}
\begin{equation}
\epsilon(H_{3}) =\mu_{B}\left\{\frac{2\pi \hbar c N}{Ae} + \frac{H^{2}_{3}}{H'_{3}}\left(\frac{H'_{3}}{H_{3}}-\left[\frac{H'_{3}}{H_{3}}\right]\right)\left(1- \frac{H'_{3}}{H_{3}}-\left[\frac{H'_{3}}{H_{3}}\right] \right)\right\},
\label{eqE:18}
\end{equation}
\end{widetext}
where $\mu_{B}$ is Bohr's magneton.
\begin{equation}
\mu_{B}= \frac{e\hbar}{2mc}.
\label{eqE:19}
\end{equation}
$M/\mu_{B}$ is obtained by deriving Eq.~\eqref{eqE:18} with respect to $H_{3}$.
\begin{eqnarray}
\frac{M}{\mu_{B}} = 1- 2\lambda -\frac{2\lambda(1-\lambda)}{\lambda+n'}.
\label{eqE:20}
\end{eqnarray}
Fig.~\ref{fig5} plots the magnetization versus $\nu=\lambda+n'$ as given by Eq.~\eqref{eqE:20}.
Notice that although the energy is a continuous function of $H_{3}$, its derivative is discontinuous when the Landau level is totally filled.
Consider the field $H_{3}$ near to a multiple of $H'_{3}$.
In this neighborhood there is a change in $\nu=\lambda+n'$ by going  from $\lambda=1^{-}$ (slightly below 1.0)  to $\lambda=0^{+}$ (slightly above 0).
A discontinuity arises in the magnetization since  $\frac{M(\lambda=1^{-})}{\mu_{B}} \approx -1.0$ and $\frac{M(\lambda=0^{+})}{\mu_{B}} \approx 1.0$, as shown in Fig.~\ref{fig5}.

In case many Landau levels are filled, namely case $n'>>1$, Eq.~\eqref{eqE:20} becomes,
\begin{eqnarray}
\frac{M}{\mu_{B}} \approx 1- 2\lambda,
\label{eqE:21}
\end{eqnarray}
which shows that the magnetization  $M$ is a periodic function of $1/H_{3}$ with the period given by,
\begin{eqnarray}
\Delta\left(\frac{1}{H_{3}} \right) = \frac{1}{H'_{3}}=\frac{A}{N\Phi_{0}}.
\label{eqE:22}
\end{eqnarray}
Using that  $ /N =4\pi/k^{2}_{F}$ (spinless fermions) we retrieve the de Haas-van Alphen result that the oscillations are inversely proportional to the
Fermi surface area $A_{F} \equiv \pi k^{2}_{F}$:
\begin{eqnarray}
\Delta\left(\frac{1}{H_{3}} \right) = \frac{2\pi e}{\hbar c}\frac{1}{A_{F}}.
\label{eqE:23}
\end{eqnarray}
%

\subsection{The attractive magnetic interaction among particles confined to the lowest Landau level}

In this section we include interaction among particles confined to the lowest Landau level.
The motion of the particles due to the external field make them generate currents which create a local magnetic field causing a mutual interaction among them.
We prove here that this magnetic interaction is attractive and to treat it the field energy is incorporated into the hamiltonian, and added to the kinetic energy such that,
\begin{equation}
\mathcal{H} = K+F,
\label{eqc:01}
\end{equation}
where the kinetic energy, $K$, is,
\begin{equation}
K = \int d^{3}x \frac{1}{2m} \sum^{2}_{j=1} \left ( D_{j}\Psi\right)^{\dag}\left ( D_{j}\Psi\right) ,
\label{eqc:02}
\end{equation}
%
and the field energy, $F$, is,
\begin{equation}
F = \int d^{3}x \frac{1}{8\pi} \left(\vec{h}(\Psi)-\vec{H} \right)^{2}.
\label{eqc:03}
\end{equation}
The local magnetic field is also a second quantized field,  $\vec{h}(\Psi)$, obtained by solving Amp\`ere's law,
\begin{eqnarray}
\hspace{-0.2cm}\vec{\nabla}\times\vec{h}(\Psi) = \frac{4\pi}{c}\vec{J}(\Psi), \!\quad \vec{J}(\Psi)=\frac{e}{2m}\!\left( \Psi^{\dag}\vec{D}\Psi + c.c. \right),
\label{eqc:04}
\end{eqnarray}
by knowledge of the current. A special attention must be paid to the dimensionality of the fields, as the magnetic field and the the current are tied to each other through Amp\`ere's law,
which assigns to  $\Psi$ the dimensionality $[\Psi]=1/\sqrt{V}$, where $V=A L_3$ is a volume, the product of the unit cell area in the plane, $A$, times a length along the direction of the applied field, $L_3$.
This is consistent with the energy being an integral over  three-dimensional space, but uniaxial symmetry along the direction of the external applied field renders the problem two-dimensional.
All the fields only depend on $(x_1,x_2)$, including the local field is $\vec{h}(\Psi) = \hat{x}_{3}h_{3}(\Psi)$,  $h_{3}=\partial_{1} A_{2}-\partial_{2} A_{1}$.
For the applied field we choose, as before, that $A_{3}=0$ and $\vec{H}=\hat{x}_{3}H_{3}$.\\
Remarkably Amp\`ere's law is solved exactly and the local magnetic field  $h_3(\Psi)$ fully determined.
To achive this purpose we  introduce a dual view of the kinetic energy that requires the operators,
\begin{equation}
D_{\pm} \equiv D_{1}\pm iD_{2}.
\label{eqc:05}
\end{equation}
We use the following identity
\begin{eqnarray}
\left ( D_{+}\Psi \right)^{\dag}\left ( D_{+}\Psi \right)  &=& \sum^{2}_{j=1}\left( D_{j}\Psi \right )^{\dag}\left( D_{j}\Psi \right ) + i\Psi^{\dag}[D_1,D_2]\Psi \nonumber \\ &-&\frac{\hbar m}{e}\left(\partial_{1}J_{2} -\partial_{2}J_{1} \right),
\label{eqc:06}
\end{eqnarray}
where the current components $J_{i}$, $i=1$ e $2$, are given by Eq.~\eqref{eqc:04}, and the commutation relation is given by,
\begin{equation}
[D_1,D_2]=-\frac{e\hbar}{ic}h_3.
\label{eqc:07}
\end{equation}
Combining  Eqs. \eqref{eqc:02}, \eqref{eqc:06}, \eqref{eqc:07}, and using that,
\begin{eqnarray}
\frac{\hbar m}{e}\left(\partial_{1}J_{2} -\partial_{2}J_{1} \right) = \frac{\hbar^{2}}{2}\vec{\nabla}^{2}(\Psi^{\dag}\Psi),
\label{eqc:08}
\end{eqnarray}
one obtains that,
\begin{eqnarray}
\hspace{-0.2cm} K &=&
\int\! d^{3}x \left( \frac{\vert D_{+}\Psi\vert^2}{2m}  + \frac{e\hbar}{2mc}h_3\Psi^{\dag}\Psi\right) \nonumber \\
&+& \frac{\hbar^{2}}{4m}\int d^3x \vec{\nabla}^{2}(\Psi^{\dag}\Psi),
\label{eqc:09}
\end{eqnarray}
where the last term is a surface term.
The dual formulation of the kinetic energy also  leads to a dual formulation of the current, which is obtained by varying the kinetic energy with respect to the vector potential,
$\delta F_k = \cdots -\frac{1}{c}\int d^2x\vec{J} \cdot \delta\vec{A}$.
Therefore one obtains the components of the current in the plane.
\begin{eqnarray}
J_{1}=\frac{e}{2m}\big[\Psi^{\dag}\big(D_{+}\Psi\big)+\big(D_{+}\Psi\big)^{\dag}\Psi\big]-\frac{e\hbar}{2m}\partial_{2}\left(\Psi^{\dag}\Psi\right),
\label{eqc:10}
\end{eqnarray}
e
\begin{eqnarray}
\hspace{-0.2cm} J_{2}=\frac{e}{2im}\big[\Psi^{\dag}\big(D_{+}\Psi\big)-\big(D_{+}\Psi\big)^{\dag}\Psi \big]+\frac{e\hbar}{2m}\partial_{1}\left(\Psi^{\dag}\Psi\right).
\label{eqc:11}
\end{eqnarray}
The confinement to the lowest Landau leads to a full solution of Amp\`ere's law. The axial symmetry gives that $\partial_{2} h_{3} = (4\pi/c)J_{1}$ e $\partial_{1} h_{3} = -(4\pi/c)J_{2}$.
Then $h_{3} + (2\pi e\hbar/mc)\vert\Psi\vert^2= \mbox{constant}$, and this constant is determined by the condition that the local field must be equal to the applied field, $h_{3}=H_{3}$, in case that  $\Psi=0$.
We have found that there are first order equations that link $\Psi$  to $h_{3}$, in the following way.
\begin{eqnarray}
D_{+}\Psi=0,
\label{eqc:12}
\end{eqnarray}
and
\begin{eqnarray}
 h_{3}(\Psi) = H_{3} - 4\pi\mu_{B}\Psi^{\dag}\Psi,
\label{eqc:13}
\end{eqnarray}
where  $\mu_{B}$ is Bohr's magneton, Eq.~\eqref{eqE:19}.
The first order equations are expected to be a useful approximation to the solution in case that  $h_3 \ll H_3$.
For this reason they are only solved approximately in the following way.
Firstly  Eq.~\eqref{eqc:12} is solved for $\Psi$ assuming that the vector potential is only due to $H_3$.
Thus corrections to vector potential are not considered in this stage.
Then the local magnetic field, $h_3$, is obtained from Eq.~\eqref{eqc:12}, which leads to corrections to $H_3$ in terms of  $\Psi$.
The general solution for $\Psi$, given by Eq.~\eqref{eqa:04}, is the $n=0$ part of the general solution, since $D_{+}\Psi_{0,p}=0$.
\begin{eqnarray}
\Psi_{0,p}(x_{1},x_{2}) =  \frac{1}{\sqrt{L_{3}}}\sum^{p-1}_{q=0}c_{0,q}\phi_{0,q}(x_{1},x_{2},p).
\label{eqc:16}
\end{eqnarray}
where
\begin{eqnarray}
\hspace{-0.1cm}\phi_{0,q}(x_{1},x_{2},p) = \!\left(\frac{2\pi p}{L^{3}_{1}L_{2}} \right)^{\frac{1}{4}}\sum^{+\infty}_{l=-\infty}\! e^{i\frac{2\pi}{L_1}(pl+q)x_{1}}e^{-\frac{1}{2}f_{lq}^{2}},
\label{eqc:17}
\end{eqnarray}
with $f_{lq}$ defined in \eqref{eq0:21} since $D_{+}\phi_{0,q}=0$.
Once in power of $\Psi$ the local field follows from \eqref{eqc:13} by computing the expectation value $\left\langle \Phi \big{\vert}h_{3}(\Psi_{0,p})\big{\vert}\Phi\right\rangle$ for a state $\vert\Phi\rangle$ constrained to the first Landau level.
Hence one obtains that,
\begin{eqnarray}
\left\langle \Phi \big{\vert} h_{3}(\Psi_{0,p})\big{\vert}\Phi\right\rangle = H_3 -\mu_{B} \frac{4\pi}{L_{3}} \sum^{p^{\prime}-1}_{q=0}\big{\vert}\phi_{0,q}(x_{1},x_{2},p) \big{\vert}^{2}.
\label{eqc:18}
\end{eqnarray}
\\

The first order equations  unveil an attraction among particles confined to the lowest Landau level.
To see this introduce Eq.~\eqref{eqc:12} into the kinetic energy, given by Eq.~\eqref{eqc:09}. The Hamiltonian \eqref{eqc:01} becomes $\mathcal{H} = \mu_{B} \int\! d^{3}x h_3\Psi^{\dag}\Psi + \frac{\hbar^{2}}{4m}\int d^{3}x \vec{\nabla}^{2}(\Psi^{\dag}\Psi) + \frac{1}{8\pi}\int d^{3}x \left(\vec{h}-\vec{H} \right)^{2}$, that once combined with $\Psi^{\dag}\Psi = -(1/4\pi\mu_{B})(h_{3} - H_{3})$ from Eq.~\eqref{eqc:13} renders that,
\begin{equation}
\hspace{-0.2cm} \mathcal{H} = \int d^{3}x\left(\frac{1}{8\pi}H_{3}^{2}-\frac{1}{8\pi}h_{3}(\Psi)^2 + \frac{\hbar^{2}}{4m}\vec{\nabla}^{2}(\Psi^{\dag}\Psi) \right),
\label{eqc:14}
\end{equation}
Next we express $\mathcal{H}$ by introducing Eq.~\eqref{eqc:13} into Eq.~\eqref{eqc:14}.
\begin{equation}
\hspace{-0.3cm} \mathcal{H} = 2\pi\mu^{2}_{B}\int d^{3}x\left[ \frac{H_{3}\Psi^{\dag}\Psi}{2\pi\mu_{B}}-\left(\Psi^{\dag}\Psi\right)^2 +\frac{\vec{\nabla}^{2}(\Psi^{\dag}\Psi)}{2\pi r_{e}} \right],
\label{eqc:15}
\end{equation}
where Bohr's magneton is related to the electron's classical radius by $\hbar^{2}/4m=\mu^{2}_{B}/r_{e} \rightarrow r_{e}=e^{2}/mc^{2}$.
Eq.~\eqref{eqc:14} shows that the magnetic field energy is negative, $-h_3^2/8\pi$, and so, able to lower the energy by becoming more intense.
In other words, it causes attraction among the particles.
Notice that the total energy must remain positive, as initially assumed according to  Eq.~\eqref{eqc:01}.
For the periodic state the last term vanishes, $\int d^{2}x \vec{\nabla}^{2} \Psi^{\dag}\Psi=0$, as previously shown.
Since $\mu_B H_3=\hbar \omega_c/2$, the Hamiltonian cal also be expressed as,

\begin{equation}
\mathcal{H} =\int d^{3}x\left[ \frac{\hbar \omega_c}{2}\Psi^{\dag}\Psi-2\pi\mu^{2}_{B}\left(\Psi^{\dag}\Psi\right)^2  \right],
\label{eqc:15b}
\end{equation}
The second term is evidently attractive although, in comparison with the first, very small.\\

The length $L_3$ is truly a free parameter that in fact determines the local magnetic field $h_3$, as shown in  Eq.~\eqref{eqc:18}.
Thus it  cannot be determined by the present approach.
It should be regarded as a phenomenological parameter that gauges the difference $h_3-H_3$.
Recall that the three dimensional integration is turned into a two dimensional integration, by assuming $\int d^{3}x/L_3 =\int d^{2}x$.
Thus this length limits the validity of the first order equation approach, which is to have $h_3 - H_3$ very small.
According to Eq.~\eqref{eqc:13} this requires that $H_{3} \gg 4\pi\mu_{B}\Psi^{\dag}\Psi$, or equally,
\begin{eqnarray}
L_3>> \sqrt{\frac{2\pi}{p}}r_e.
\end{eqnarray}
This is easily reachable for any reasonable $L_3$, considering that $r_e= 2.8 \; 10^{-15}$ m.
This relation is obtained by assuming  that $\Psi^{\dag}\Psi \sim \sqrt{2\pi p}/(L_3 A)$, where the area $A$ is associated to the flux quantization in the plane, $H_3 A=p \Phi_0$.
Bohr's magneton, the magnetic flux and the classical radius are connected through $\mu_B= \frac{\Phi_0 r_e}{4\pi}$, and from it the above condition for $L_3$ is straightforwardly obtained.
In summary the deviation $H_3-h_3$ is  freely adjusted and once this is done the parameter $L_3$ becomes known.
A part from the choice of $L_3$  the theory can be treated two-dimensionally in the plane $(x_1,x_2)$ due to its uniaxial symmetry along the applied field.\\

\begin{figure}[t!]
\centering
\includegraphics[scale=0.48]{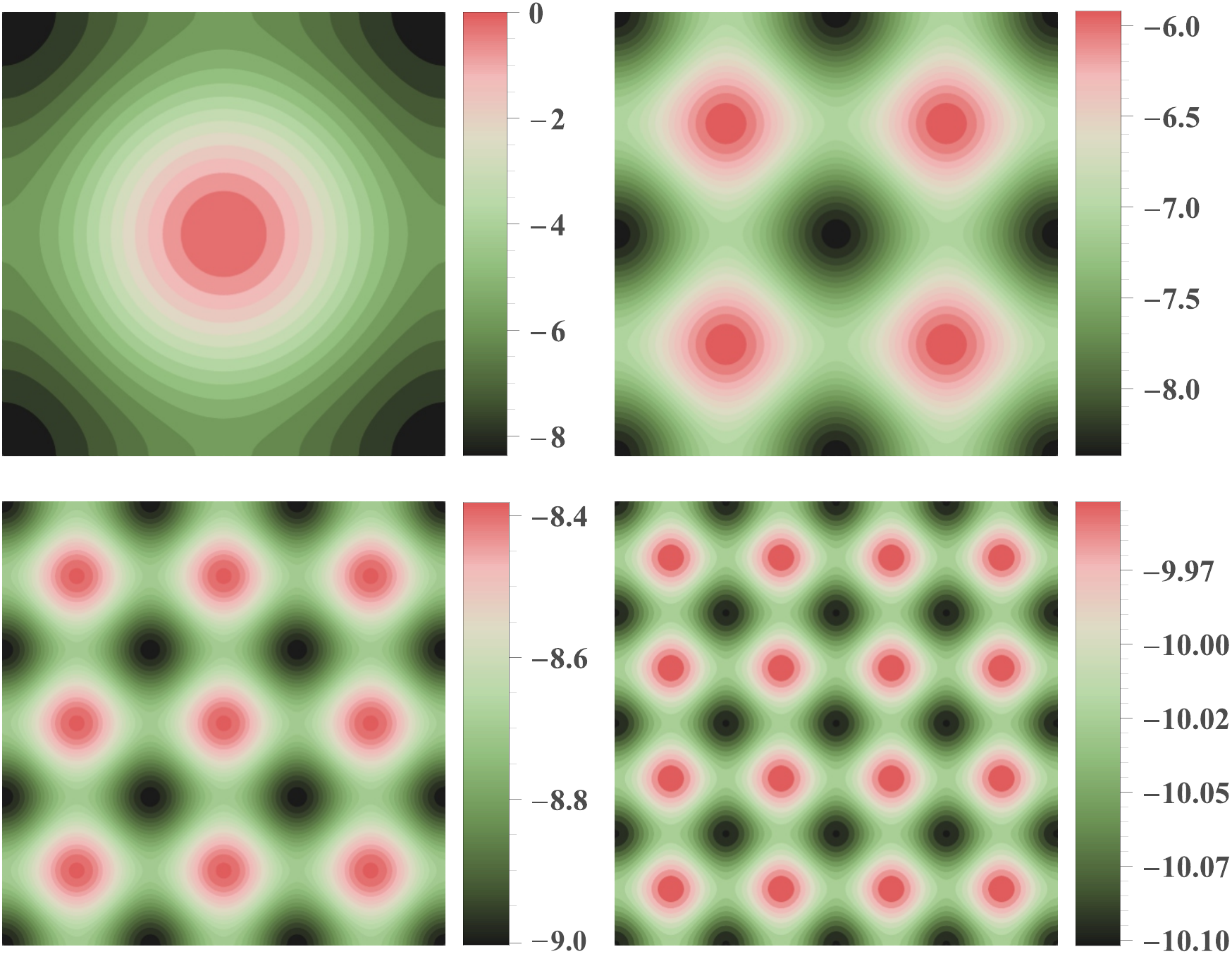}
\caption{The deviation of the local magnetic field to the applied field, $-\left\langle \Phi \big{\vert} h_{3}(\Psi_{0,p})-H_3\big{\vert}\Phi\right\rangle L_{3}/\mu_{B}$, is shown here in a unit cell.
The unit cell $(x_{1}/L_{1}, x_{2}/L_{2})$ is displayed for the cases $p=1$ (left,up), $p=2$ (right,up), $p=3$ (left,down) and $p=4$ (right,down ), respectively.
The number of particles for the totally filled lowest Landau level, $n=0$ is $N=p$.
Thus there are $1$ (top-left), $2$ (top-right), $3$ (bottom-left), and $4$ (bottom-right) particles (fluxons), respectively.
Remarkably the ratio between the number of maxima (minima) of the local field, as seen in the above plots,  and the number of particles is $1$, $1/2$, $1/3$ and $1/4$, respectively. }
\label{fig6}
\end{figure}

The deviation of the local field to the applied field is shown in Fig.~\ref{fig6}, and is given by   $-\left\langle \Phi \big{\vert} h_{3}(\Psi_{0,p})-H_3\big{\vert}\Phi\right\rangle L_{3}/\mu_{B}$.
This deviation shows the existence of maxima and minima  in the unit cell, $(x_{1}/L_{1},x_{2}/L_{2})$.
The lowest Landau level is assumed totally filled and the figure shows the cases $p=1$ (left,up), $p=2$ (right,up), $p=3$ (left,down) and $p=4$ (right,down).
Notice that while this figure shows the density, Fig.~\ref{fig6} displays the local field $h_3$, and they are connected to each other by  Eq.~\eqref{eqc:18}.
An egg-box pattern emerges with  $p^2$ maxima (pink) and $p^2$ minima (black) in the magnetic field pattern.
The ratio between the maxima (minima) of the local field and the number of particles is fractional, and given by $1/p$, as found in Table~\ref{tab-ratio}.

%
\section{Conclusion}

The set of  orthonormal bases developed here yield wave functions for the well-known Schr\"odinger equation problem of a particle in a magnetic field that features a periodic density probability.
These wave functions display a fixed number of $p$ trapped magnetic flux per Landau level.
The second quantization study reveals that the number of particles, $N$,  and the number of fluxons trapped in the unit cell, $p$, are independent.
For $N$ fermions distributed among $n+1$  Landau energy levels we find the remarkable property  that the density of particles presents $p^2$ spatial maxima (minima) in the unit cell for any $n$.
In case the highest Landau level is completely filled, and so with $N=(n+1)p$  particles, there are $(n+1)/p$ particles per maxima (minima).
We have shown that in case that particles fall in the lowest Landau level ($n=0$), the magnetic field produced by the particles yields a residual attractive interaction among them.
The present set of orthonormal functions retrieves the de Haas-van Alphen effect results, namely the total magnetization displays a periodicity with respect to the inverse of the applied field which is proportional to the area of the Fermi surface.

%
\bibliography{reference1}
\end{document}